\documentclass{nature}

\usepackage{soul}
\usepackage{color, colortbl}
\usepackage{makecell}
\usepackage{url}
\usepackage{graphicx}
\usepackage{adjustbox}
\usepackage{multirow} 
\usepackage{listings}
\usepackage[hidelinks]{hyperref}
\usepackage{cleveref}

\usepackage{float}

\usepackage{xcolor}

\title{Mobility constraints in segregation models}

\author{Daniele Gambetta$^{1,2}$, Giovanni Mauro$^{1,2,3}$ \& Luca Pappalardo$^1$
{
\scriptsize 
\\ \texttt{ daniele.gambetta@phd.unipi.it}, \texttt{giovanni.mauro@phd.unipi.it}, \texttt{luca.pappalardo@isti.cnr.it}
}
}

\begin{document}

\maketitle

\begin{affiliations}
 \item Institute of Information Science and Technologies, National Research Council (ISTI-CNR).
 \item University of Pisa, Italy.
 \item IMT School for Advanced Studies, Lucca, Italy.
\end{affiliations}

\begin{abstract}
Since the development of the original Schelling model of urban segregation, several enhancements have been proposed, but none have considered the impact of mobility constraints on model dynamics. 
Recent studies have shown that human mobility follows specific patterns, such as a preference for short distances and dense locations. This paper proposes a segregation model incorporating mobility constraints to make agents select their location based on distance and location relevance.
Our findings indicate that the mobility-constrained model produces lower segregation levels but takes longer to converge than the original Schelling model. 
We identified a few persistently unhappy agents from the minority group who cause this prolonged convergence time and lower segregation level as they move around the grid centre. 
Our study presents a more realistic representation of how agents move in urban areas and provides a novel and insightful approach to analyzing the impact of mobility constraints on segregation models. 
We highlight the significance of incorporating mobility constraints when policymakers design interventions to address urban segregation.
\end{abstract}

\section{Introduction}

In 1971, economist Thomas Schelling proposed an agent-based model to explain how individual actions could result in global phenomena, focusing on urban segregation \cite{schelling1971dynamic, schelling1969models, hegselmann2017thomas, schelling2006micromotives}. 
He observed that segregation dynamics emerge due to homophily among social groups across various demographic factors such as ethnicity, language, income, and class affiliation \cite{schelling2006micromotives}.
To illustrate this idea, Schelling used a simple spatial proximity model that divided the population into two groups based on a homophily threshold. 
Agents of two colours were placed randomly on a two-dimensional grid, and each agent had a preference for living next to people in their group. If an agent is unhappy with their current location, they will move to the nearest square that satisfies them. 
Schelling found that segregation emerges above a homophily threshold of 1/3, and other factors affecting segregation include the ratio of individuals, the homophily threshold, and individual demands.

Numerous variants and enhancements of the Schelling model have been proposed so far, modifying agents' behaviour \cite{laurie2003role, mantzaris2020incorporating, sert2020segregation}, environmental configuration \cite{fossett2009effects, fagiolo2007segregation, freeman1978segregation, henry2011emergence, rogers2011unified, vinkovic2006physical}, considering geographical regions \cite{gimblett2002integrating,benenson2004geosimulation,benenson2002entity}, including real-world segregation data along with strategies to validate simulated behaviour with observations \cite{park2018high,venkatramanan2018using,zhang2016data}, implementing agent behaviours based on psychological and sociological theories \cite{scalco2018application,wang2015abm,schrieks2021integrating,abella2022aging}, and allowing for sensitivity analysis to quantify outcome dependency on various parameters and initial conditions \cite{borgonovo2022sensitivity, ligmann2020one, niida2019sensitivity, clark2008understanding}.
Other works show how even milder preferences or integration policies can eventually lead to unexpected segregation scenarios \cite{zhang2011tipping, zhang2004residential}, and how the introduction of venues can have an impact on segregation dynamics \cite{silver2021venues}.

Despite these advancements, all proposed models assume that unhappy agents move randomly on the grid without any preference for nearby or far away locations. 
However, recent empirical studies have shown that human movement, far from being random, follows specific statistical patterns across various spatial scales, including daily movements and migrations \cite{gonzalez2008understanding, pappalardo2013understanding, pappalardo2015returners, barbosa2018human, alessandretti2020scales, schlapfer2021universal, song2010modelling, simini2012universal, simini2021deep, hu2018life, zhao2016unified, hu2019return, luca2021survey, bohm2022gross}.
These individual mobility patterns are characterized by a preference for short distances and relevant places over longer distances and sparse ones \cite{gonzalez2008understanding, pappalardo2013understanding, pappalardo2015returners, barbosa2018human, alessandretti2020scales, schlapfer2021universal, zipf2016human, zipf1946p, simini2021deep, mauro2022generating, luca2022modeling}.
Despite considerable interest in modelling and predicting human mobility \cite{luca2021survey, barbosa2018human}, it remains unclear how mobility patterns relate to segregation patterns.
Recent empirical studies suggest a link between experienced income segregation and an individual's tendency to explore new places and visitors from different income groups \cite{moro2021mobility}, but this has not been systematically studied in an agent-based, Schelling-like simulation framework.
At the same time, although a few simulation studies show how restricting relocation options \cite{fagiolo2007segregation} and considering collective factors \cite{abella2022aging,grauwin2009competition} affect segregation dynamics, these findings have not been related to mobility constraints. 
Thus, we still lack a comprehensive understanding of how mobility constraints impact segregation dynamics. 

This study fills this gap by designing a segregation model that considers mobility constraints and exploring how they influence segregation dynamics. 
Drawing on the gravity law of human mobility \cite{zipf, zipf1946p, simini2021deep, liu2020universal, prieto2018gravity, lenormand2016systematic}, our model allows unhappy agents to select the next location to move based on distance and location relevance.
Our findings reveal that mobility-constrained models exhibit lower levels of segregation than the Schelling model, albeit with a longer convergence time, and that agents that end in the periphery are more segregated than those in the grid centre.
We attribute these phenomena to a small group of persistently unhappy agents from the minority group who gravitate towards the grid centre due to the preference for nearby and relevant locations imposed by mobility constraints.

\section{Mobility-constrained segregation models}

In our model, agents may be of two types, moving on a bi-dimensional grid of size $m = N \times N$. 
As in the original Schelling model \cite{schelling1971dynamic, schelling1969models, hegselmann2017thomas, schelling2006micromotives}, an agent is happy when it is surrounded by a number of agents of the same type above a predetermined homophily threshold.
An unhappy agent located at a cell $A$ moves to a new cell $B$ based on a probability function, $p(B)$, which depends on two factors: the distance $d(A, B)$ between $A$ and $B$, and the relevance $r(B)$ of destination $B$.  
This probability captures the gravity law of human mobility \cite{zipf, zipf1946p, simini2012universal, simini2021deep, liu2020universal, prieto2018gravity,lenormand2016systematic, barbosa2018human, luca2021survey,pappalardo2022scikit}, positing that people tend to travel to nearby and relevant locations, a concept that has been supported by extensive research in fields ranging from transport planning \cite{erlander1990gravity} and spatial economics \cite{prieto2018gravity, karemera2000gravity, patuelli2007network} to epidemic spreading \cite{luca2022modeling, balcan2010modeling, li2011validation, cevik2022going, zhang2020exploring}.
The distance between points $A$ and $B$, represented by coordinates $(x_A, y_A)$ and $(x_B, y_B)$, is computed as their Euclidean distance on the grid, $d(A, B)=\sqrt{(x_A-x_B)^2 + (y_A-y_B)^2}$.
Mathematically, we define the probability of an agent moving to cell $B$, given its current cell $A$, as a product of two power-law functions: 
\begin{equation}
p(B) \propto r(B)^\alpha d(A, B)^\beta
\end{equation}
where parameter $\alpha > 0$ models the tendency to move preferably to relevant places, while $\beta$ captures the tendency to prefer ($\beta > 0$) or avoid ($\beta <0$) large displacements.
We assume a core-periphery structure to model the distribution of relevance across the grid cells \cite{louf2016patterns} and use a radial distribution where the relevance value of each cell decreases with its distance from the grid centre $C$: 
\begin{equation}
r(A) \propto \frac{1}{d(A,C)^\kappa}
\end{equation}
with $\kappa=2$.
The results obtained with a uniformly random spatial distribution of relevance can be found in Supplementary Note 1. 
Note that since all agents share the information about cell relevance, $\alpha = -x$ means being repelled by a cell to the same extent that $\alpha = x$ means being attracted to it.
The case where $\alpha = 0$ and $\beta = 0$ corresponds to the original Schelling model.
The model simulation ends when all agents are happy.
From the gravity segregation model, we derive two other families of models: the distance models ($\alpha=0, \beta \neq 0$), which only imposes constraints on distance, and the relevance models ($\alpha \neq 0, \beta = 0$), which only considers relevance.
See Methods for algorithmic details about the gravity, relevance, and distance models.

We conduct a series of experiments varying the grid size ($m \in \{25 \times 25, 50 \times 50, 75 \times 75\}$), the ratio of occupied cells ($\sigma \in \{50\%, 70\%\}$), the proportion of agents in the two groups ($\theta = \{ 10/90, 30/70, 50/50\}$), the homophily threshold ($h \in \{10\%, 30\%, 50\%\}$), and the relocation policy (whether agents move to locations where they should be happy or not). 
Since our results are consistent across these different parameter values (see Supplementary Note 2), we present the findings for a fixed set of conditions: a grid size of $50\times 50$, an occupancy ratio of $70\%$, a group proportion of $30/70$, and a homophily threshold of $30\%$. 
We perform 100 simulations for each set of parameter values, each time using a different random distribution of agents on the grid.
Each simulation terminates when all agents are happy or after a maximum of $500$ simulation steps. 

We quantify the level of segregation at the end of the simulation as the average segregation level of the agents, $S = \frac{\sum_{a \in M}s(a)}{|M|}  $, where $M$ is the set of agents and $s(a)=\frac{|\Gamma_{T}(a)|}{|\Gamma(a)|}$, with $\Gamma_{T}(a)$ the set of neighbours of the same type of $a$ and $\Gamma(a)$ the set of neighbours of $a$ of any type.
$\Gamma_{T}(a)$ is computed as the Moore neighbourhood with range $r=1$ (see Methods).
We obtain analogous results when using other segregation metrics (see Supplementary Note 3).

\section{Results}

\begin{figure}
    \centering
\includegraphics[width=\textwidth]{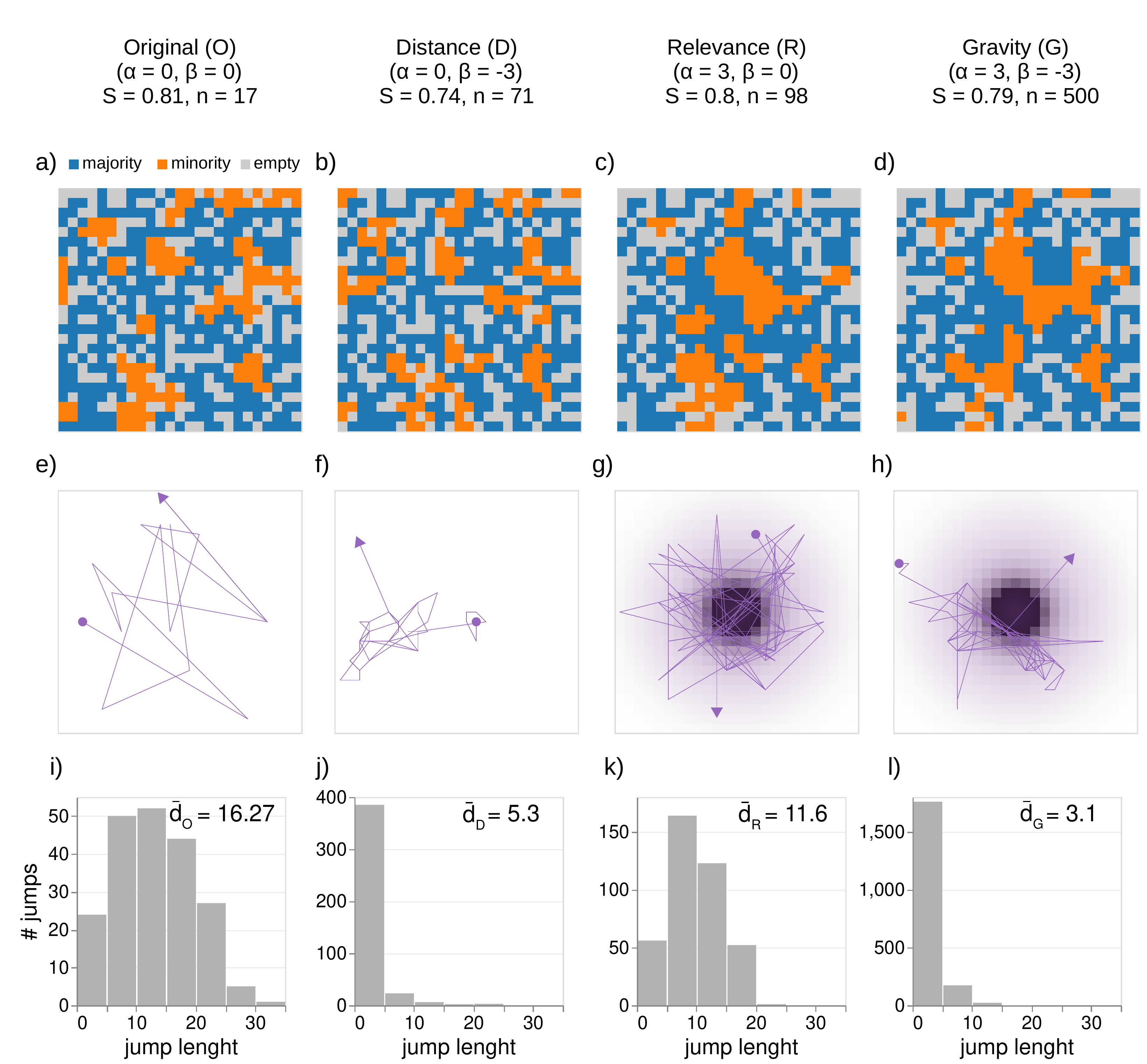}
    \caption{\footnotesize \textbf{Segregation dynamics of mobility-constrained models.}
    (a-d) Final grid configuration of four simulations of the original Schelling model $O$ (a), a distance model $D$ (b), a relevance model $R$ (c), and a gravity model $G$ (d), all with a grid size of $25 \times 25$, a cell occupancy rate of $70\%$, a group proportion of $30/70$, and a homophily threshold of $30\%$. 
    (e-h) The mobility trajectories of four agents, one per model. 
    (i-l) Graphical representation of the distribution of jump length for each model.
    In $O$, agents move randomly throughout the grid when they are unhappy, resulting in a peaked distribution of jump lengths. 
    In $D$ and $G$, agents prefer to move to nearby cells, resulting in a heavy-tailed distribution of jump lengths. 
    In $R$ and $G$, agents are directed towards the city centre, resulting in a centripetal tendency that causes empty cells to be positioned far away from the centre and a lower average distance travelled than $O$. 
    We indicate the average distance travelled by agents in the four models with $\overline{d}_O$, $\overline{d}_D$, $\overline{d}_R$, and $\overline{d}_G$.}
    \label{figure1}
\end{figure}

The values of parameters $\alpha$ and $\beta$ influence how agents move on the grid when unhappy, generating different mobility patterns. 
As an example, in Figure \ref{figure1}, we compare three simulations with identical initial configuration but different parameter values: $\alpha = \beta= 0$, representing the original Schelling model $O$; $\alpha=0, \beta=-2$, a distance model $D$; $\alpha=2, \beta=0$, a relevance model $R$; and $\alpha=2, \beta=-2$, a gravity model $G$.
In the original Schelling model ($O$), agents move randomly throughout the grid when they are unhappy, resulting in a distribution of jump lengths that follows a peaked distribution, indicating the existence of a typical jump length for agents ($\overline{d_{O}} = 16.27$, the average distance between cells on the grid). 
On the other hand, in the distance model ($D$) and the gravity model ($G$), agents prefer to move to nearby cells, resulting in a heavy-tailed distribution of jump lengths (Figure \ref{figure1}j, k) and lower average distances ($\overline{d_{D}} = 5.30$ and $\overline{d_{G}} = 3.10$). 
In $R$ and $G$, agents are directed towards the city centre, resulting in a lower average distance travelled ($\overline{d_{R}}=11.60$) than $O$ and a centripetal tendency that causes the empty cells to be positioned far away from the centre of the grid (Figure \ref{figure1}c, d).
Indeed, in $R$ and $G$, only $4\%$ and $7\%$ of the cells in the centre are empty, while in $D$ and $O$, $25\%$ and $28\%$ of cells are empty, respectively.
An intriguing question is what impact $\alpha$ and $\beta$, and thus the resulting mobility patterns, have on crucial aspects of the simulation, such as the final segregation level, $S$, and model convergence time, $n$, defined as the number of simulation steps needed for all the agents to become happy. 


\subsection{Effects of distance on segregation dynamics.}
\label{sec:beta}

We study the impact of varying the distance exponent $\beta$ on the final level of segregation $S$ and the convergence time $n$ while holding the relevance parameter constant ($\alpha=0$). 

\begin{figure}
    \centering    \includegraphics[width=0.925\textwidth]{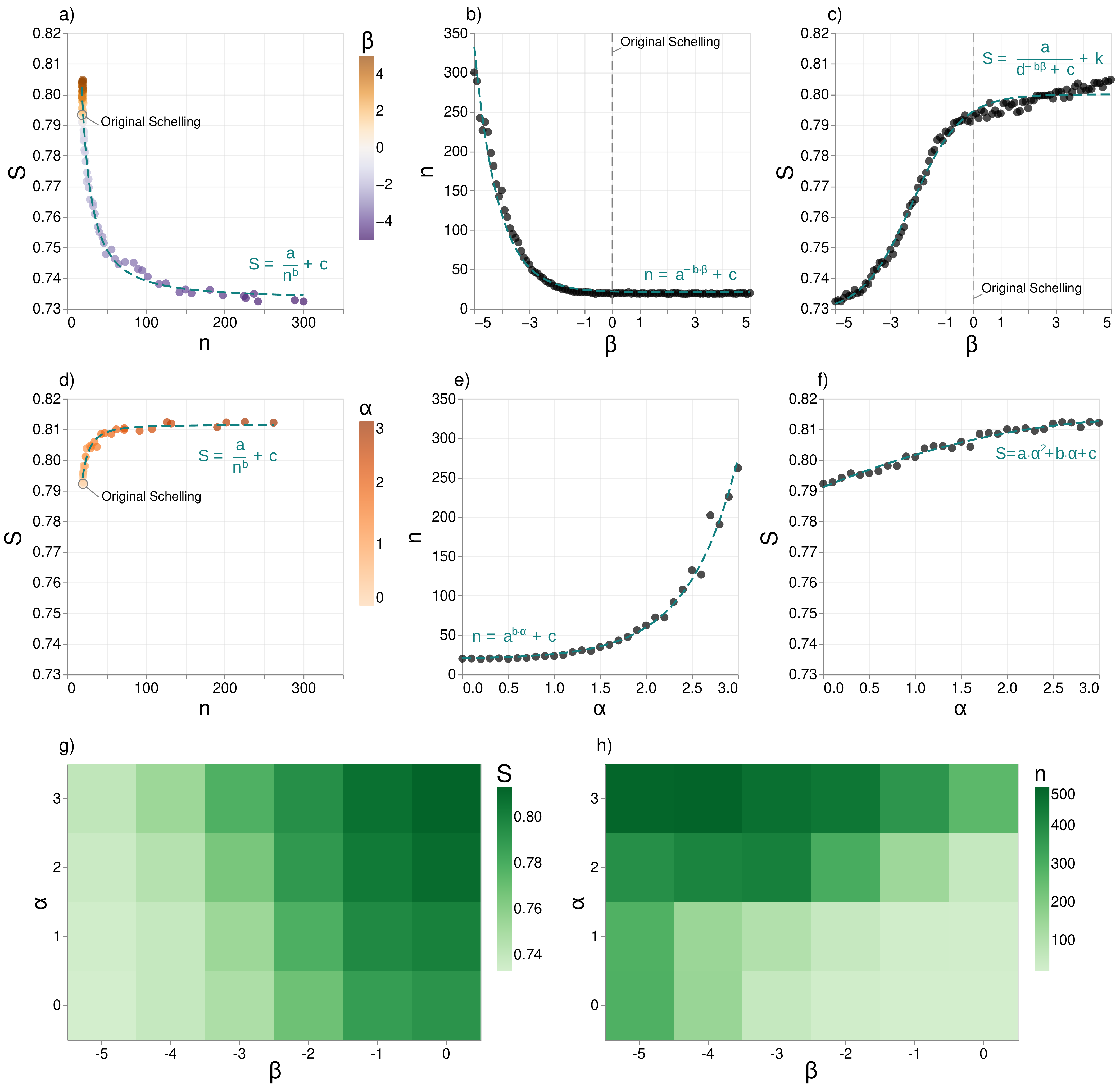}
    \caption{\scriptsize \textbf{Effects of distance and relevance on segregation dynamics.} 
    (a-c) Effects of $\beta$ on segregation dynamics. (a) The average value of $n$ and $S$ over 100 simulations with the same $\beta$ value but different initial grid configurations, colour-coded by the value of $\beta$. 
    The lower $\beta$, the higher the cost of relocating far away, resulting in longer convergence time and reduced segregation levels compared to the original Schelling model. 
    The relationship between $n$ and $S$ is well fitted by a power-law function $S = \frac{a}{n^b}+c $, with $a=4.6 , b=1.5 , c=0.7$.
    (b) $\beta$ vs average $n$ over 100 simulations. 
    The relationship follows an exponential distribution $n = a^{-b \beta}+c$, with $a = 2.7 , b = 1.1 , c = 21.0$.
    The lower $\beta$ ($< 0$), the longer the simulation.
    (c) $\beta$ vs average $S$ over 100 simulations.
    This is well approximated by a sigmoid function $S = \frac{a}{d^{-b\beta}+c}+k$, with $a = 0.9 , b = 3.2 , c = 11.8 , d = 1.4 , k = 0.7$
    For $\beta < 0$, there is an exponential increase in $S$; $\beta > 0$, the growth is moderate.
    (d-f) Effects of $\alpha$ on segregation dynamics. 
    (d) The average value of $n$ and $S$ over 100 simulations with the same value of $\alpha$ but different initial grid configurations, colour-coded by the value of $\alpha$.
    Increasing values of $\alpha$ elongate $n$ and slightly increase $S$. 
    The relationship follows a power-law function $S = \frac{a}{n^b}+c$, with $a = -8.5 , b = 2.1 , c = 0.8$.
    (e) $\alpha$ vs average $n$ over 100 simulations. 
    The relationship follows an exponential distribution $n = a^{b\alpha}+c$, with $a = 7.1 , b = 0.9 , c = 19.5$.
    (f) $\alpha$ vs average $S$ over 100 simulations.
    The relationship is well approximated by a parabolic function $S = a\alpha^2 +b\alpha+c$, with $a = -0.002, b = 0.01, c = 0.8$.
    (g) The average $S$ (colour) for each combination of $\alpha$ and $\beta < 0$. 
    For every value of $\alpha$, higher $\beta$ values lead to a higher $S$; for every $\beta$, higher $\alpha$ values lead to a higher $S$.
    (h) The average $n$ (colour) for each combination of $\alpha$ and $\beta < 0$. 
    For every value of $\alpha$, higher $\beta$ values lead to a lower $n$; for every $\beta$, higher $\alpha$ values lead to higher $n$.
    }
    \label{figure2}
\end{figure}

Figure \ref{figure2}a displays the relationship between $n$ and $S$ for $\beta \in [-5, 5]$.  
We find that a decrease in $\beta$ (i.e., an increasing cost of relocating far away) leads to a decrease in $S$ and a increase in $n$ compared to the original Schelling model ($\beta = 0$).
This result means that a preference to move nearby the current location slows down but mitigates the segregation behaviour.
For instance, for $\beta=-5$, $S$ is reduced by around $10\%$, and $n$ increases by a factor of 30 compared to the original Schelling model ($\beta=0$).
Conversely, an increase in $\beta$ (lower cost of relocating far) has a negligible impact on $n$ and marginally increases $S$. 
Indeed, for $\beta = 5$, the segregation level increases only by around $1\%$ compared to the original Schelling model, with no substantial change in the convergence time.
As an example, Figure \ref{figure1}a-d presents the final grid configurations for $O$, $R$, $D$, and $G$: note how $D$ and $G$ (with $\beta = -2$), which converge over a longer time, have a lower final segregation level compared to $O$ and $R$.

We also observe that $\beta$ influences $n$ and $S$ in a non-linear fashion.
The dependency of $n$ and $\beta$, can be approximated by an exponential function:
\begin{equation}
n = a^{-b \beta}+c
\label{eq:steps_beta}
\end{equation}
with $a = 2.7 , b = 1.1 , c = 21$ (Figure \ref{figure2}b).
Equation \ref{eq:steps_beta} enables us to estimate the number of simulation steps required for the model to converge based solely on the value of $\beta$, providing valuable what-if insights into the implications of mobility constraints on segregation dynamics. For example, if an incentive for relocating far away from the current location is introduced, which would result in a positive value of $\beta$ (e.g., $\beta=5$), Equation \ref{eq:steps_beta} suggests that it would take approximately $n=21$ simulation steps for the city to become segregated. On the other hand, if people are incentivised to stay nearby the current location, which would result in a negative value of $\beta$ (e.g., $\beta=-5$), Equation \ref{eq:steps_beta} suggests that an average of $n=256$ simulation steps (more than ten times more) would be required for the city to become segregated. 

The relationship between $\beta$ and $S$ is well-fitted by a sigmoid function:
\begin{equation}
S = \frac{a}{d^{-b\beta}+c}+k
\label{eq:seg_beta}
\end{equation}
where $a = 8.5 \cdot 10^{-1} , b = 3.2 , c = 11.8 , d = 1.4 , k = 7.3 \cdot 10^{-1}$ have been empirically fitted (Figure \ref{figure2}c).
An interesting finding is a tipping point in the sigmoid curve close to $\beta=0$ (original Schelling model), which indicates that the influence of $\beta$ on the level of segregation depends on whether it is positive or negative (Figure \ref{figure2}c).
When $\beta > 0$, agents move far way over a larger pool of choices because the number of available cells increases with distance, leading to behaviour that closely resembles the original Schelling model (see Figure \ref{figure2}a).
In fact, even for $\beta=5$, the final segregation level $S$ is only $1\%$ higher than the original Schelling model (Figure \ref{figure2}c).
In contrast, when $\beta < 0$, agents prefer to stay close to their current cells, causing the grid configuration to change slowly over time and leading to a final grid configuration that is more similar to the initial one (and therefore less segregated).
For example, if $\beta=-5$, the segregation level $S=0.73$, about 8\% lower than the original Schelling model.

The relation between $S$ and $n$ is well-fitted by a power-law function: 
\begin{equation}
S = \frac{a}{n^b}+c
\label{eq:seg_steps}
\end{equation}
with $a=4.6$, $b=1.5$, and $c=0.7$ (Figure \ref{figure2}a).
As $n$ increases, the overall level of segregation $S$ decreases: the more difficult it is for the agents to find cells that meet their homophily preferences, the less segregated the final grid is, and the more steps are required to reach an equilibrium state where all agents are happy. Conversely, if the agents can quickly identify desirable locations, convergence occurs rapidly with fewer steps, resulting in greater segregation. 

\subsection{Effects of relevance on segregation dynamics.}
\label{sec:alpha}

We study the impact of the relevance exponent $\alpha$ on the final level of segregation $S$ and the convergence time $n$ by controlling for the distance parameter ($\beta=0$). 
We find that $n$ increases exponentially with $\alpha$ as:
\begin{equation}
n = a^{b \cdot \alpha} + c 
\end{equation}
where $a = 7.1$, $b = 0.94$, and $c=19.5$ are empirically fitted (Figure \ref{figure2}e).
This result suggests that agents tend to move closer to the opposing group when they compete for relevant cells, resulting in a more prolonged convergence process than the original Schelling model.
For example, $\alpha=3$ corresponds to an average of 261 simulation steps, almost 12 times more than the needed for $\alpha=0$ (20 steps on average).
On the other hand, $S$ increases parabolically with $\alpha$ as:
\begin{equation}
    S = a \cdot \alpha^2 + b \cdot \alpha + c
\end{equation}
where $a = -0.002$, $b=0.01$, $c=0.8$ empirically fitted (Figure \ref{figure2}f), indicating a slightly non-linear impact of $\alpha$ on the final segregation level.
When $\alpha=3$, $S=0.81$, a mere $2.5\%$ higher than the original Schelling model ($0.79$), indicating that the relevance parameter has only a minor effect on the final segregation level.

The relationship between $n$ and $S$ through the relevance exponent $\alpha$ can be approximated by power-law function:
\begin{equation}
S = \frac{a}{n^b}+c
\end{equation}
with $a = -8.5, b = 2.1, c = 0.8$  (Figure \ref{figure2}d).
The longer it takes for the model to converge (larger $n$), the more segregated the final grid becomes (higher $S$, albeit only slightly). 
We observe that as $\alpha$ increases, the difference in segregation levels between the periphery and centre, denoted as $S_{\tiny \mbox{diff}} = S_{\tiny \mbox{periphery}} - S_{\tiny \mbox{centre}}$, also increases. 
This means that agents that end up in the periphery are more segregated than those in the centre (see Supplementary Note 4). 
This can be attributed to the centripetal force that encourages agents to move towards the centre, creating a more mixed scenario than in the periphery. 
Since the periphery is larger and shows a higher segregation level than the centre, the final grid becomes more segregated (higher $S$) as $\alpha$ increases.

\subsection{Interplay of distance and relevance.}

Finally, we study the interplay of the relevance and distance exponents on segregation dynamics. 
We focus on the negative values of $\beta$, which induce the most significant variations in $S$ and $n$. For completeness, we provide an analysis of $\beta>0$ in Supplementary Note 5. 

Segregation is maximized in the relevance model ($\beta = 0$) and minimized when $\beta=-5$, with a minor influence of parameter $\alpha$ (Figure \ref{figure2}g).
Specifically, $S_{\beta=0, \alpha=3}$ is approximately 11\% higher than $S_{\beta=-5, \alpha=0}$.
This suggests that when the preference for short distances is strong, the impact of relevance is relatively weak.
Figure \ref{figure2}h shows that the convergence time is maximized when $\beta$ is low and $\alpha$ is high (upper left corner). 
In particular, $n_{\beta=-5, \alpha=3}$ is roughly 24 times higher than $n_{\beta=0, \alpha=0}$.
Distance constraints become almost irrelevant when relevant places are present in influencing convergence time. 
Overall, our results highlight the complexity of the interplay between $\beta$ and $\alpha$ and their differential impacts on segregation level and convergence time.

\begin{figure}
    \centering    \includegraphics[width=.9\textwidth]{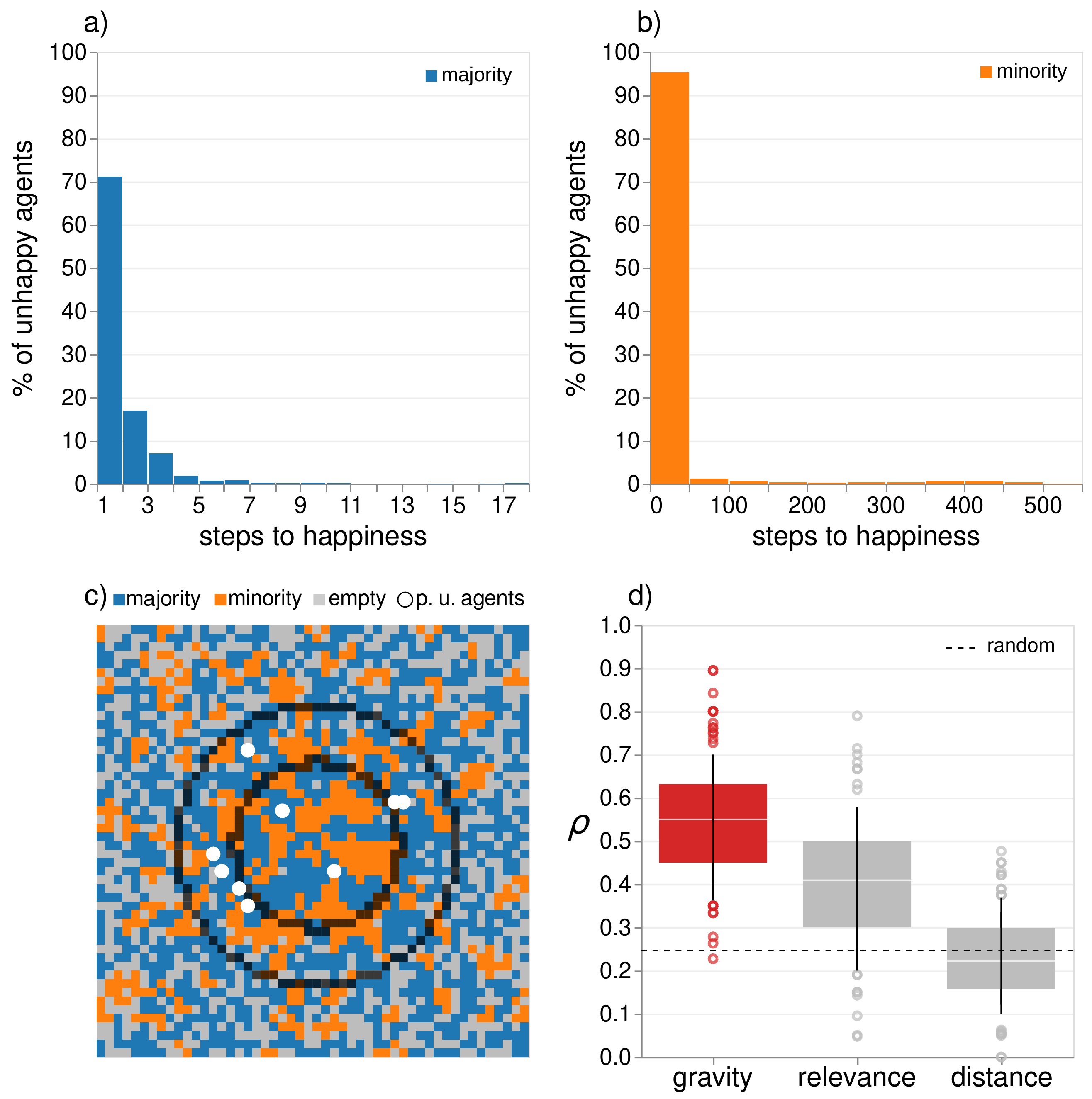}
    \caption{\footnotesize \textbf{Persistently unhappy agents.} 
    (a) Distribution of simulation steps required for the agents in the majority group to achieve happiness (100 simulations of the gravity model, $\alpha=3, \beta=-3$). 
    All agents reach happiness within 17 steps. 
    (b) Distribution of simulation steps needed for agents in the minority group to reach happiness. 
    Most agents become happy in 50 steps, but a small percentage remain unhappy for up to 500 steps. 
    (c) Gravity model's grid configuration at the critical point when the centre becomes stably segregated. 
    The boundaries of the suburbia are represented as two dark circles with a radius of 9 (small circle) and 16.6 (large circle), representing 25\% of the grid's total area. 
    The white dots represent persistently unhappy agents who remain unhappy for more than the 95th percentile of simulation steps. 
    Most of these agents are in the suburbia region. 
    (d) Probability $\rho$ of a persistently unhappy agent being in suburbia at the critical point when the centre becomes stably segregated. 
    The dashed line indicates the probability of a random agent being in the area (0.25). 
    The gravity and relevance models show higher probabilities than the random model, while the distance model's probability is similar to the null model.}
    \label{figure3}
\end{figure}

One intriguing aspect is the identification of agents responsible for elongating convergence time in the gravity models. 
Do all agents suffer prolonged unhappiness, or is there a specific group of agents responsible for the extended convergence time?
Figure \ref{figure3}a-b illustrates the distribution of time-to-happiness, representing the number of simulation steps needed for agents to become happy. 
Our results indicate that most agents achieve happiness within a few steps, while a small fraction of agents, primarily from the minority group, remain unhappy for an extended duration.
For example, as Figure \ref{figure3}a shows, all agents of the majority group achieve happiness within 17 simulation steps. 
In contrast, although most minority group agents attain happiness relatively quickly, some remain unhappy for an extended period for up to 500 simulation steps (Figure \ref{figure3}b).
We classify these agents as persistently unhappy and denote them as set $U$. 
Specifically, we identify persistently unhappy agents as those whose unhappiness surpasses the 95th percentile of the distribution of time-to-happiness for all agents.
For example, in Figure \ref{figure3}c, only 9 agents out of  1750 are persistently unhappy.

We hypothesize that the combination of agents' preferences and the distribution of relevant cells within the grid creates a situation where some agents may be trapped in an area that does not align with their preferences, resulting in persistent unhappiness.
We discover that this area is a ``suburbia'' of the centre, i.e., an area around the centre between a radius of 9 and a radius of 16.6, representing 25\% of the grid's total area (see Figure \ref{figure3}c).
In detail, we compute the fraction $\rho$ of persistently unhappy agents being located in the suburbia at a critical point when the centre becomes stably segregated (see Supplementary Note 6).  
For instance, Figure \ref{figure3}c illustrates the grid configuration of a gravity model at this critical point: the majority of persistently unhappy agents (white dots) are concentrated in a confined region near the centre of the grid.
Mathematically, we capture this aspect by defining $\rho = \frac{|U_{sub}|}{|U|}$, where $|U_{sub}|$ is the number of persistently unhappy agents in the suburbia at the critical point.
Figure \ref{figure3}d shows that the gravity and the relevance models have a significantly higher $\rho$ than a random spatial distribution of agents, with the relevance model displaying a lower probability than the gravity model. 
The distance model’s probability is instead similar to the null model.
These findings suggest that persistently unhappy agents are located in suburbia at the critical point, and their presence can significantly prolong the simulation as they cannot escape the loosely segregated situation in the centre due to their preference for relevant locations and short-distance moves.

\section{Discussion} 

This study provides a novel and insightful approach to understanding the impact of mobility constraints on urban segregation dynamics. 
The gravity law injected in the Schelling model allows for modelling agents' mobility patterns based on two factors: the relevance of locations and the distance between them. 
We find that the influence of these two factors on segregation dynamics is significant and profoundly impacts the model outcome, both independently or in combination with each other. 
By analyzing the mobility patterns of agents in the gravity-constrained model, we discovered a trend in the exponential elongation of segregating times attributed to a few agents of the minority group caught in a vicious loop, constantly moving among the few locations near the segregated centre.
Moreover, the centripetal force towards the centre leads to more mixing in the centre and greater segregation in the periphery.

Our findings provide valuable insights into the complex dynamics of segregation, offering a valuable tool for understanding and simulating potential scenarios through what-if analysis. 
The equations that relate the relevance and distance parameters with the segregation level and the convergence time can be manipulated to explore different scenarios, such as economic incentives to encourage large-scale relocations or the relocation of facilities within a city. This approach can provide policymakers with a comprehensive understanding of how these interventions may impact segregation dynamics.

Although this study relies on a simplified representation of a city, our future research aims to replicate the analysis on a real-world dataset of relocations to assess how simulation-based results align with empirical observations in an actual city. 
This approach can have a significant impact on the comprehension of urban dynamics and the understanding of the factors that drive segregation in real-world scenarios.

\section{Methods}
\label{sec:methods}

\subsection{Segregation models}

Given the input parameters, which are grid size $m$, the ratio of occupied cells $\sigma$, the proportion of agents in the two groups $\theta$, homophily threshold $h$, relevance exponent $\alpha$ and distance exponent $\beta$, the simulation of the segregation models is executed based on the following algorithm.

\paragraph{Model Initialization.} To initialize the model, agents are randomly assigned to one of two groups based on the parameter $\theta$. The agents are then placed randomly on a two-dimensional grid with density defined by $\sigma$. In the original and distance models, each cell on the grid is of equal relevance. 
The relevance and gravity models assign relevance to each cell using $r(A) \propto \frac{1}{d(A, C)^2}$, where $C$ indicates the centre of the grid.

\paragraph{Simulation Step.} At each simulation step, agents are randomly activated and evaluate their happiness based on the number of same-group neighbours they have: an agent is happy if the fraction of neighbours of the same type is higher or equal to $h$. 
The neighbourhood of a cell $A = (x_A, y_A)$ on the grid is computed as its Moore neighborhood \cite{gardner1970fantastic}, $\mathcal{N}_{(x_A,y_A)}=\{(x,y):|x-x_A| \leq r,|y-y_B| \leq r \}$, with $r=1$. 
If an agent is unhappy, it will move to an empty cell on the grid. 
In the original Schelling model, the empty cell is selected randomly with uniform probability. The relevance model chooses the empty cell $B$ based on probability $P(B) = r(B)^\alpha$. 
In the distance model, the empty cell $B$ is chosen based on probability $d(A, B)^\beta$.
In the gravity model, the selection is based on the joint probability function $P(B) = r(B)^\alpha d(A, B)^\beta$.

\paragraph{Model Termination.} The simulation ends when all the agents are happy, or a maximum of 500 simulation steps are reached.







\begin{addendum}
 \item This work has been partially supported by 1) EU project H2020 Humane-AI-net G.A. 952026; 2) EU project H2020 SoBigData++ G.A. 871042; 3) NextGenerationEU - National Recovery and Resilience Plan (Piano Nazionale di Ripresa e Resilienza
 , PNRR), project “SoBigData.it - Strengthening the Italian RI for Social Mining and Big Data Analytics”, prot. IR0000013, avviso n. 3264 on 28/12/2021.
 We thank Daniele Fadda for his precious support on data visualization.

\item[Code Availability]
Code availability
The Python code to replicate the models and all the experiments is available at \url{https://github.com/dgambit/mobility_schelling}.

\item[Data Availability]
All data generated during this study are included in this published article and its supplementary information files and can be found in the code repository available at \url{https://github.com/dgambit/mobility_schelling}.
 
 \item[Competing Interests] The authors declare that they have no
competing financial interests.
\item[Authors contribution] GM and LP conceived and conceptualized the research. DG made the simulations, the data analysis, and developed the code. DG, GM, and LP conceived the figures. DG made the plots. GM, and LP wrote the paper. LP directed and supervised the research. All authors contributed to the scientific discussion, read and approved the paper.
 \item[Correspondence] Correspondence and requests for materials
should be addressed to D.G., G.M., and L.P~(email: daniele.gambetta@phd.unipi.it, giovanni.mauro@phd.unipi.it, luca.pappalardo@isti.cnr.it).
\end{addendum}

\section*{References}
\bibliographystyle{naturemag}
\bibliography{biblio.bib}

\clearpage

\renewcommand{\thefigure}{S\arabic{figure}}

\setcounter{figure}{0}

\newpage
\section*{\Large{Supplementary Notes}}

\subsection{Supplementary Note 1: Random spatial distribution of relevance.}

\begin{figure}[H]
    \centering
\includegraphics[width=.9\textwidth]{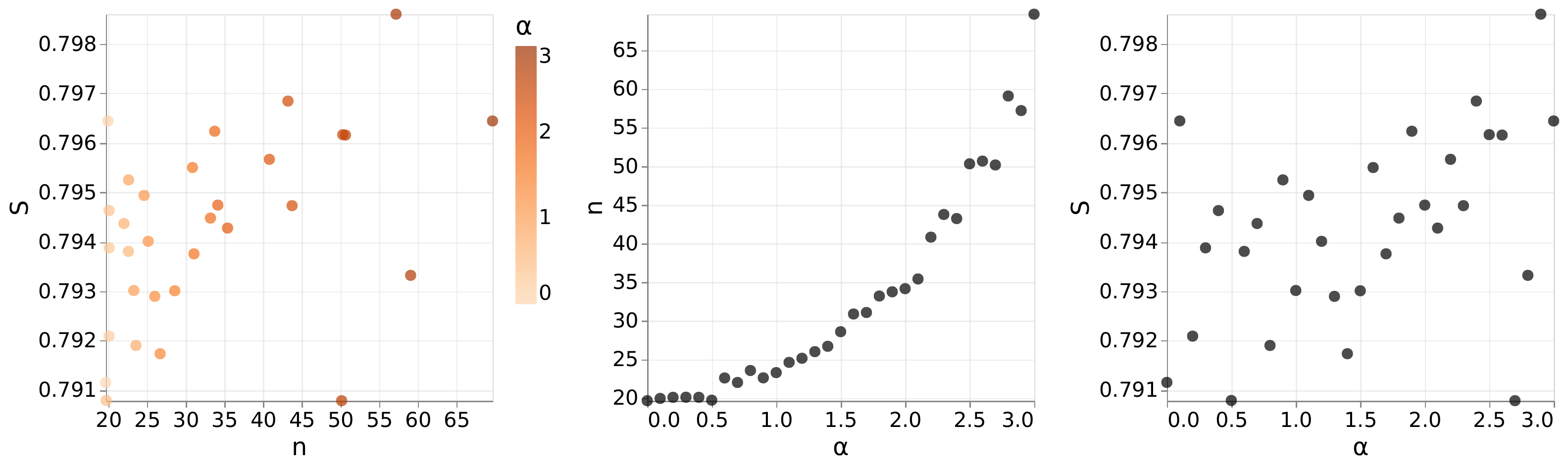}
    \caption{a) The average value of $n$ and $S$ over 100 simulations with the same value of $\alpha$ but different initial grid configurations, colour-coded by the value of $\alpha$, considering a random distribution of relevance.
    Increasing values of $\alpha$ elongate $n$ and slightly increase $S$
    (b) $\alpha$ vs average $n$ over 100 simulations. 
    (c) $\alpha$ vs average $S$ over 100 simulations.
    }
    \label{fig:random_alpha}
\end{figure}

We also explore model versions that incorporate a uniformly random spatial relevance distribution. However, as shown in Figure \ref{fig:random_alpha}, the influence of $\alpha$ on the convergence time $n$ is less prominent in this case. We still observe an exponential relationship, but the maximum value of $n$ is reduced to 70 when $\alpha = 3$, compared to 260 when assuming a core-periphery relevance distribution. Similarly, the impact on the segregation level $S$ is barely noticeable, with a minimal upward trend. In summary, adopting a uniformly random spatial distribution of relevance sacrifices realism and leads to less intriguing emerging patterns.

\subsection{Supplementary Note 2: Configurations of the mobility-constrained models.}
In addition to the configuration discussed in the manuscript, we conducted additional experiments by varying the initial model setups (grid size, majority-minority ratio, homophily percentage, relocation policy) and using employing a different agent movement policy.

\begin{figure}[H]
    \centering
\includegraphics[width=.9\textwidth]{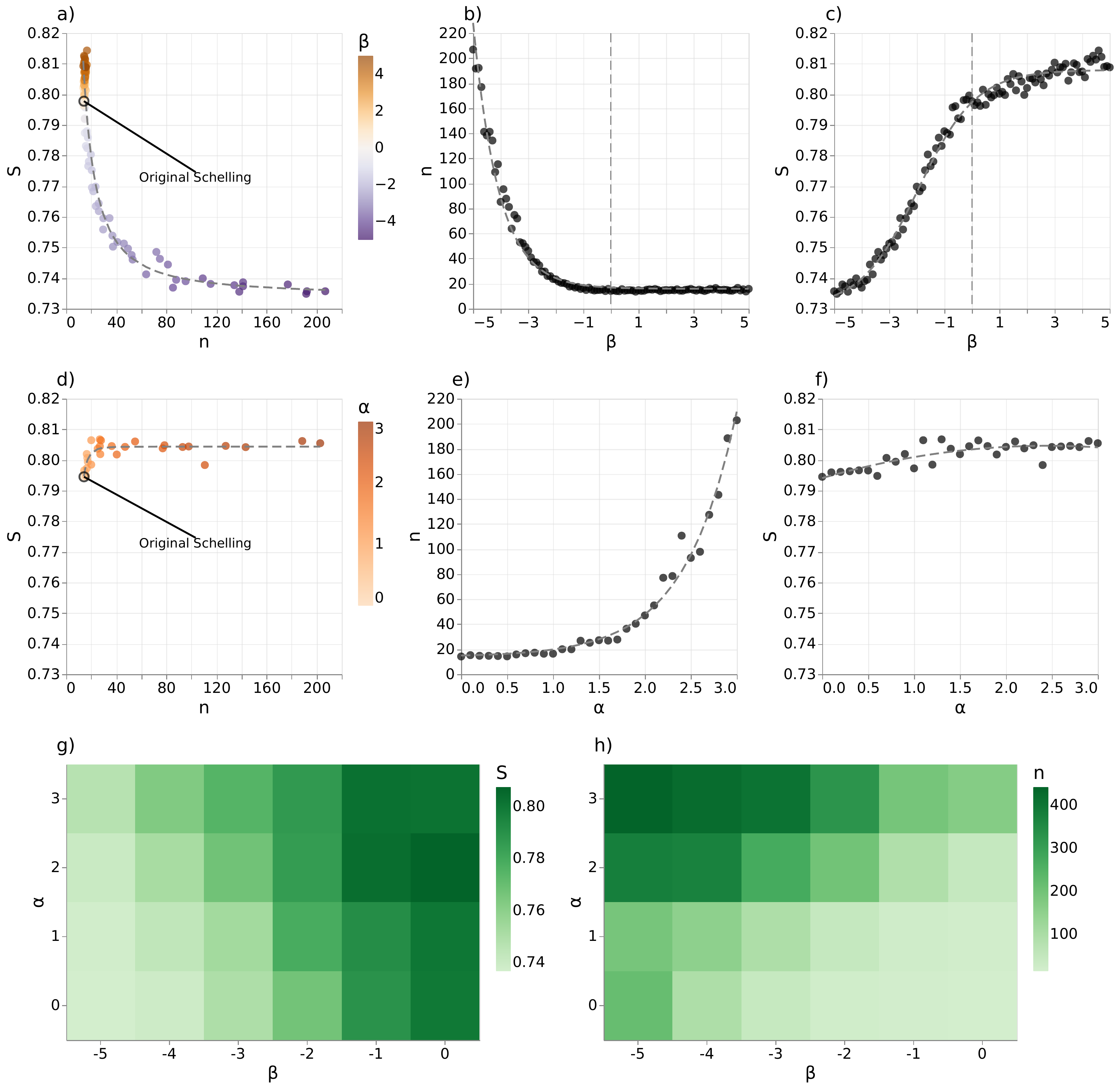}
   \caption{\scriptsize \textbf{Effects of distance and relevance exponents on segregation dynamics on a 25x25 grid} 
     (a-c) Effects of $\beta$ on segregation dynamics. (a) The average value of $n$ and $S$ over 100 simulations with the same $\beta$ value but different initial grid configurations, colour-coded by the value of $\beta$. 
    The lower $\beta$, the higher the cost of relocating far away, resulting in longer convergence time and reduced segregation levels compared to the original Schelling model. 
    (b) $\beta$ vs average $n$ over 100 simulations. 
    The lower $\beta$ ($< 0$), the longer the simulation.
    (c) $\beta$ vs average $S$ over 100 simulations.
    For $\beta < 0$, there is an exponential increase in $S$; $\beta > 0$, the growth is moderate.
    (d-f) Effects of $\alpha$ on segregation dynamics. 
    (d) The average value of $n$ and $S$ over 100 simulations with the same value of $\alpha$ but different initial grid configurations, colour-coded by the value of $\alpha$.
    Increasing values of $\alpha$ elongate $n$ and slightly increase $S$. 
    (e) $\alpha$ vs average $n$ over 100 simulations. 
    (f) $\alpha$ vs average $S$ over 100 simulations.
    (g) The average $S$ (colour) for each combination of $\alpha$ and $\beta < 0$. 
    For every value of $\alpha$, higher $\beta$ values lead to a higher $S$; for every $\beta$, higher $\alpha$ values lead to a higher $S$.
    (h) The average $n$ (colour) for each combination of $\alpha$ and $\beta < 0$. 
    For every value of $\alpha$, higher $\beta$ values lead to a lower $n$; for every $\beta$, higher $\alpha$ values lead to higher $n$.
    }    
    \label{fig:side_25}
\end{figure}

{\bf Grid size.} We conduct experiments on two additional grid configurations: a smaller grid size of $25 \times 25$ and a larger grid size of $75 \times 75$. 
Figure \ref{fig:side_25} illustrates the impact of distance and relevance exponents on segregation dynamics. 
The findings align with those presented in the manuscript, although in the smaller grid, the correlations between variables are relatively weaker compared to the larger grids.

\begin{figure}[H]
    \centering
\includegraphics[width=.9\textwidth]{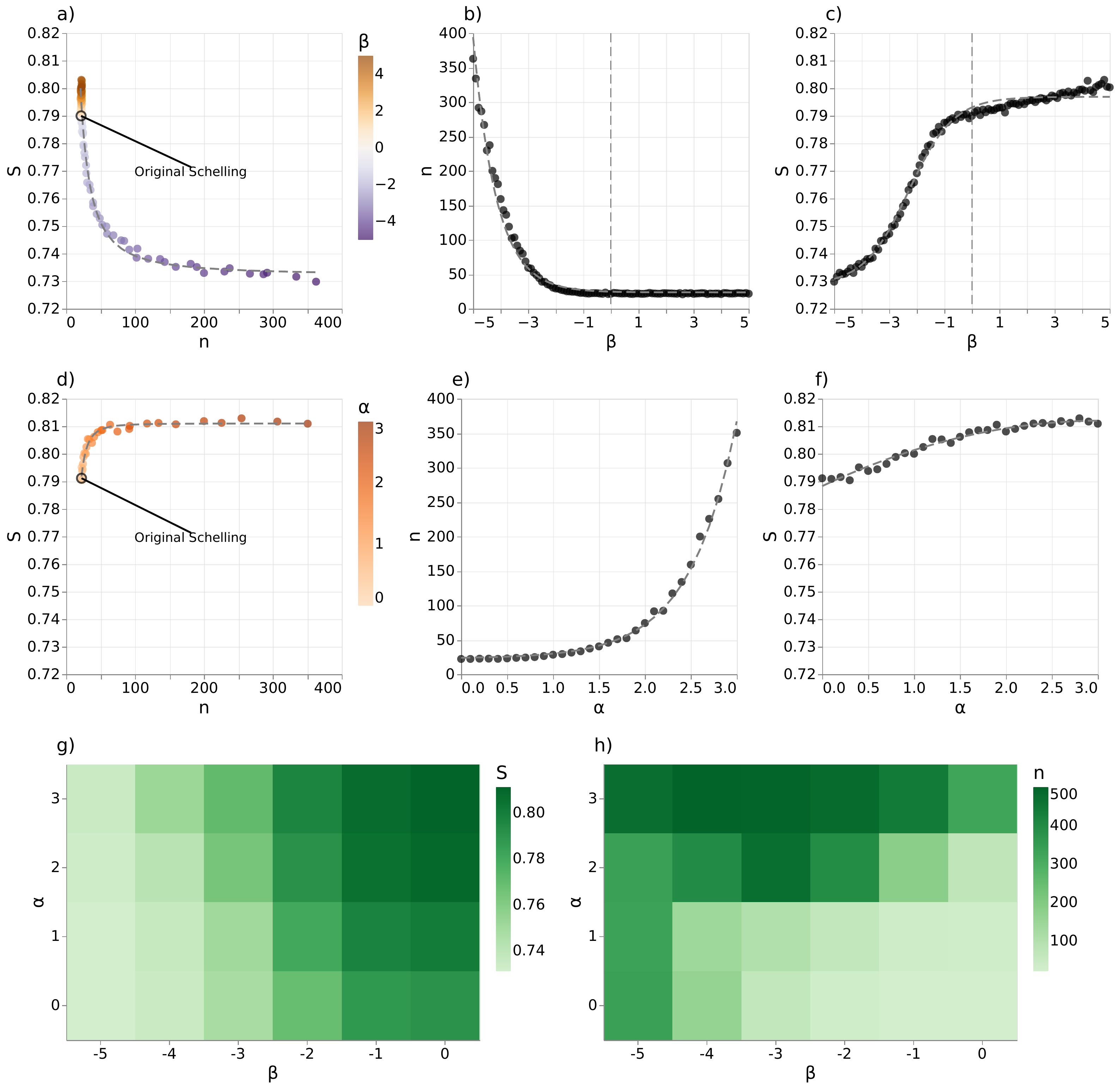}
    \caption{\scriptsize \textbf{Effects of distance and relevance exponents on segregation dynamics on a 75x75 grid} 
     (a-c) Effects of $\beta$ on segregation dynamics. (a) The average value of $n$ and $S$ over 100 simulations with the same $\beta$ value but different initial grid configurations, colour-coded by the value of $\beta$. 
    The lower $\beta$, the higher the cost of relocating far away, resulting in longer convergence time and reduced segregation levels compared to the original Schelling model. 
    (b) $\beta$ vs average $n$ over 100 simulations. 
    The lower $\beta$ ($< 0$), the longer the simulation.
    (c) $\beta$ vs average $S$ over 100 simulations.
    For $\beta < 0$, there is an exponential increase in $S$; $\beta > 0$, the growth is moderate.
    (d-f) Effects of $\alpha$ on segregation dynamics. 
    (d) The average value of $n$ and $S$ over 100 simulations with the same value of $\alpha$ but different initial grid configurations, colour-coded by the value of $\alpha$.
    Increasing values of $\alpha$ elongate $n$ and slightly increase $S$. 
    (e) $\alpha$ vs average $n$ over 100 simulations. 
    (f) $\alpha$ vs average $S$ over 100 simulations.
    (g) The average $S$ (colour) for each combination of $\alpha$ and $\beta < 0$. 
    For every value of $\alpha$, higher $\beta$ values lead to a higher $S$; for every $\beta$, higher $\alpha$ values lead to a higher $S$.
    (h) The average $n$ (colour) for each combination of $\alpha$ and $\beta < 0$. 
    For every value of $\alpha$, higher $\beta$ values lead to a lower $n$; for every $\beta$, higher $\alpha$ values lead to higher $n$.
    }
        \label{fig:side_75}

\end{figure}

{\bf Homophily.} We also investigate the impact of different levels of homophily ($0.1$ and $0.5$) on segregation dynamics, in comparison to the homophily level discussed in the manuscript. 
Figure \ref{fig:side_75} presents the results of the experiments examining the effects of distance and relevance exponents ($\beta$ and $\alpha$).
Regarding the influence of $\beta$, both cases align with the findings in the manuscript, showing similar conclusions.
In the relevance model, when the homophily level is set to $0.5$, the same conclusions as the manuscript are obtained. However, when the homophily level is reduced to $0.1$, the convergence time $n$ becomes low for all values of $\alpha$ (less than 5). 
As a result, the variation of $n$ with changing $\alpha$ becomes insignificant. This outcome can be attributed to the fact that with low homophily, agents can easily find cells where they are happy, leading to faster convergence.

{\bf Population distribution.}
We examined the impact of different population distributions between minority and majority agents: a less even distribution (10\%/90\%) and a more balanced distribution (50\%/50\%).
In the case of the 10\%/90\% distribution, we also set the density to 0.5. This adjustment was necessary because having a very high majority population led to persistent unhappiness in the minority, preventing the models from converging within 500 steps. Consequently, the analysis would have been less informative.
Figure \ref{fig:pop_10-90} an \ref{fig:pop_50-50} displays the results of the experiments investigating the effects of distance and relevance exponents on segregation dynamics.
Regarding the influence of $\beta$, both population distribution cases yielded the same conclusions as the manuscript.
In the relevance model, when the distribution was 50\%/50\%, the same conclusions as the manuscript were observed. However, when the distribution was 10\%/90\%, the convergence time $n$ was very low for all values of $\alpha$ (less than 5). As a result, the variation of $n$ with changing $\alpha$ became insignificant. This can be attributed to the fact that in the 10\%/90\% distribution, agents easily found cells where they were happy, leading to faster convergence.

{\bf Relocation policy.}
In the experiments conducted in the manuscript, we implement a movement policy where, at each step, unhappy agents moved to a random empty cell. In this section, we repeat the experiment with a modified movement policy. 
Instead of selecting a random empty cell, we only consider empty cells where an agent would be happy. 
Figure \ref{fig:good} presents the results of these experiments, illustrating the effects of distance and relevance exponents on segregation dynamics using the improved movement policy. 
In general, although the results are in line with those in the manuscript, the convergence time $n$ observed is lower compared to the manuscript because agents have a higher probability of reaching a happy configuration.

\subsection{Supplementary Note 3: Segregation metrics}
In the manuscript, we employed a classic index to measure segregation. This index was calculated by determining the average number of similar neighbors and dividing it by the total number of neighbors. An increasing value of this index indicates a greater level of segregation. In order to delve deeper into the dynamics of segregation, we conducted supplementary experiments employing an alternative measure known as the generalized Freeman Segregation Index (FSI). The FSI takes into account the presence of cross-links, which represent connections between nodes or agents belonging to different categories or types.
Here, we compute the number of cross-links, denoted as $X_n$, at each step. This quantity is obtained by summing the number of contacts between agents of different types in immediate proximity. Subsequently, we calculate the FSI at each step, denoted as $FSI_n$, by dividing the count of cross-links at that step by the count of cross-links at the beginning of the simulation (step 0). Mathematically, it can be expressed as:
$$FSI_{n} =\frac{|X_{n}|}{|X_{0}|}$$
It is worth noting that in this case, as the simulation's degree of segregation increases, the FSI values, which initially start at 1 in the first step, will decrease. This occurs because the number of neighbors of a different type diminishes as segregation intensifies.
Figure \ref{fig:fsi} demonstrates that these experiments yield the same conclusions as those presented in the manuscript. The observed trends (reversed in nature, given that an higher FSI signifies a less segregated scenario) and patterns align with the findings discussed earlier, further validating the results and reinforcing the main conclusions of the study.

\subsection{Supplementary Note 4: Segregation in the centre and the periphery}

As mentioned in the manuscript, increasing the parameter $\alpha$ tends to amplify the difference in segregation levels between the periphery and the center, as represented by $S_{\tiny \mbox{diff}} = S_{\tiny \mbox{periphery}} - S_{\tiny \mbox{centre}}$. This indicates that agents in the periphery exhibit a higher degree of segregation compared to those in the center, resulting in a more segregated final grid (refer to Figure \ref{fig:diff_seg}). Notably, when $\alpha$ reaches its maximum value ($\alpha = 3$), and $\beta$ has a significant impact ($\beta \in [-5,-3]$), the highest $S_{\tiny \mbox{diff}}$ is observed. It is noteworthy that the difference becomes almost negligible when $\alpha = 0$.

\subsection{Supplementary Note 5: Interplay of distance and relevance with $\beta>0$}

In the manuscript, we focused on the case where $\beta<0$, representing the cost of long travel for unhappy agents, and examined the segregation dynamics that arise in conjunction with the relevance parameter. However, for the sake of completeness, we also analyze the case when $\beta>0$, which assigns higher probabilities to cells farther away from the location of unhappy agents.
Figure \ref{fig:betapos} presents heatmaps illustrating the values of $S$ and $n$ for different $\beta$ values ranging from -5 to 5. In the case of positive $\beta$ values, we do not observe any significant evidence of a relationship between $S$ and $\alpha$. The segregation level, as represented by $S$, does not exhibit distinct patterns or trends corresponding to varying values of $\alpha$. However, we do find that the convergence time $n$ remains consistently low for positive $\beta$ values. Unlike the case of negative $\beta$, where $n$ varied significantly with different $\alpha$ values, the impact of $\alpha$ on $n$ is not pronounced when $\beta>0$. Despite the low convergence time, we do not observe substantial variations in $n$ across different $\alpha$ values in this scenario.
These findings highlight the contrasting dynamics between positive and negative values of $\beta$ and underscore the role of the distance parameter in shaping segregation patterns.
    
\subsection{Supplementary Note 6: Critical point of centre segregation}

As highlighted in the manuscript, the simultaneous presence of significant values for both the relevance and distance exponents results in a notable increase in the convergence time $n$. This configuration gives rise to an interesting phenomenon in the segregation dynamics of the centre, resembling a phase transition, as shown in Figure \ref{fig:tippingpoint} (with the same parameter configuration as in the manuscript). 
Specifically, the figure displays an initial sharp shift, indicating a transition from low to high levels of segregation in the centre. 
This transition is followed by an oscillation phase where the segregation value of the centre fluctuates around $0.68$. The red line represents the tipping point, which is identified as the step at which the increase in segregation value no longer exceeds 2\% compared to the previous five steps.
To gain further insights into the elongation of the convergence time $n$, we examine the distribution of unhappiness steps for each agent. We observe that the elongation is primarily caused by a few minority agents that remain unhappy for an extended period. These agents, referred to as persistently unhappy (p.u.) agents, are identified based on exceeding the 95th percentile in the distribution of steps to happiness. Remarkably, we find that these p.u. agents are predominantly located in an area known as suburbia during the tipping point step mentioned earlier. This observation holds a higher probability compared to other models, shedding light on the specific spatial patterns contributing to the prolonged convergence time in the simulation.

\clearpage

\section{Supplementary Figures}

\begin{figure}[H]
    \centering
\includegraphics[width=.9\textwidth]{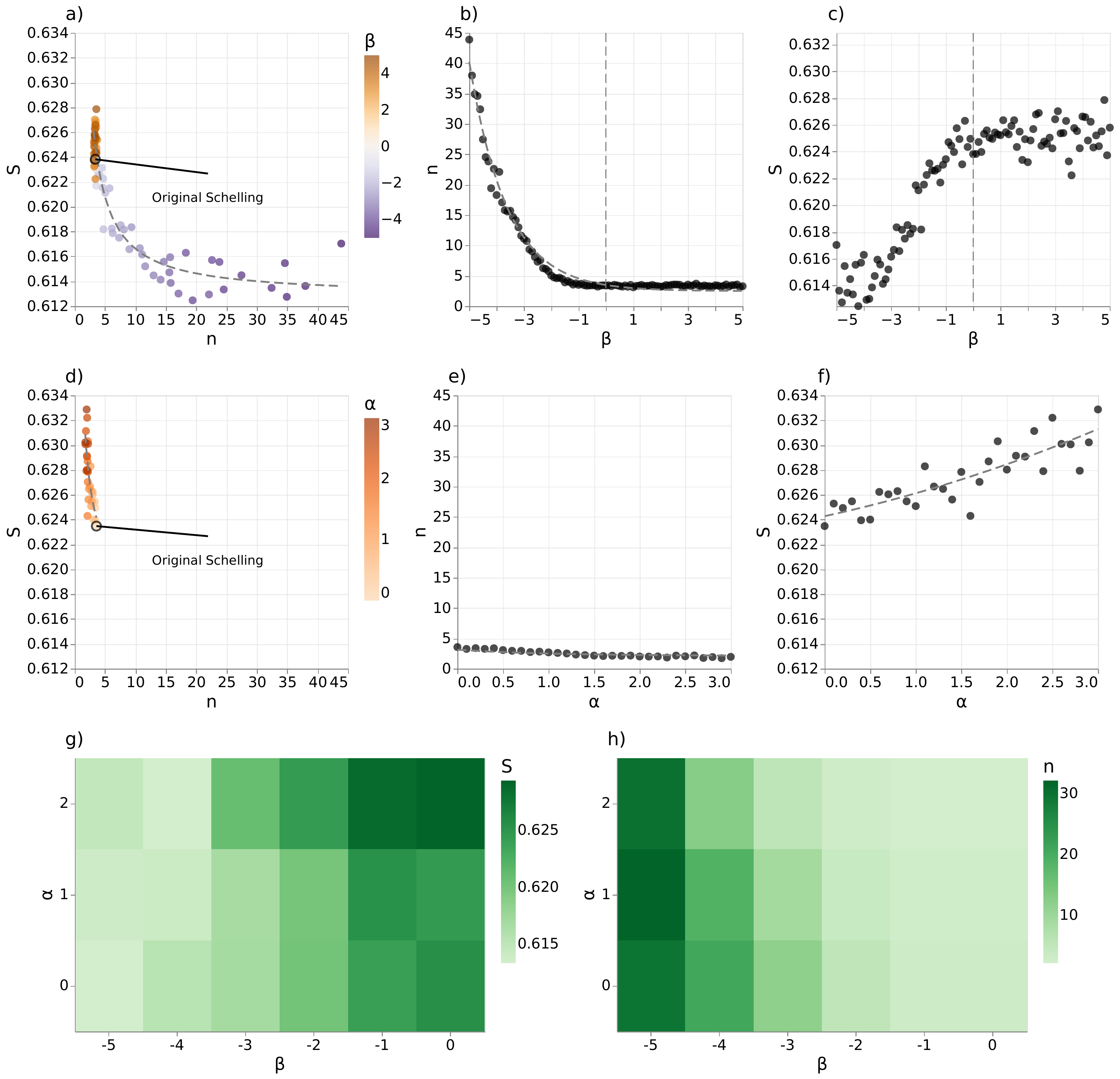}
    \caption{\scriptsize \textbf{Effects of distance and relevance exponents on segregation dynamics with homophily=0.1} 
     (a-c) Effects of $\beta$ on segregation dynamics. (a) The average value of $n$ and $S$ over 100 simulations with the same $\beta$ value but different initial grid configurations, colour-coded by the value of $\beta$. 
    The lower $\beta$, the higher the cost of relocating far away, resulting in longer convergence time and reduced segregation levels compared to the original Schelling model. 
    (b) $\beta$ vs average $n$ over 100 simulations. 
    The lower $\beta$ ($< 0$), the longer the simulation.
    (c) $\beta$ vs average $S$ over 100 simulations.
    For $\beta < 0$, there is an exponential increase in $S$; $\beta > 0$, the growth is moderate.
    (d-f) Effects of $\alpha$ on segregation dynamics. 
    (d) The average value of $n$ and $S$ over 100 simulations with the same value of $\alpha$ but different initial grid configurations, colour-coded by the value of $\alpha$.
    Increasing values of $\alpha$ elongate $n$ and slightly increase $S$. 
    (e) $\alpha$ vs average $n$ over 100 simulations. 
    (f) $\alpha$ vs average $S$ over 100 simulations.
    (g) The average $S$ (colour) for each combination of $\alpha$ and $\beta < 0$. 
    For every value of $\alpha$, higher $\beta$ values lead to a higher $S$; for every $\beta$, higher $\alpha$ values lead to a higher $S$.
    (h) The average $n$ (colour) for each combination of $\alpha$ and $\beta < 0$. 
    For every value of $\alpha$, higher $\beta$ values lead to a lower $n$; for every $\beta$, higher $\alpha$ values lead to higher $n$.
    }
        \label{fig:hom_0.1}

\end{figure}

\begin{figure}[H]
    \centering
\includegraphics[width=.9\textwidth]{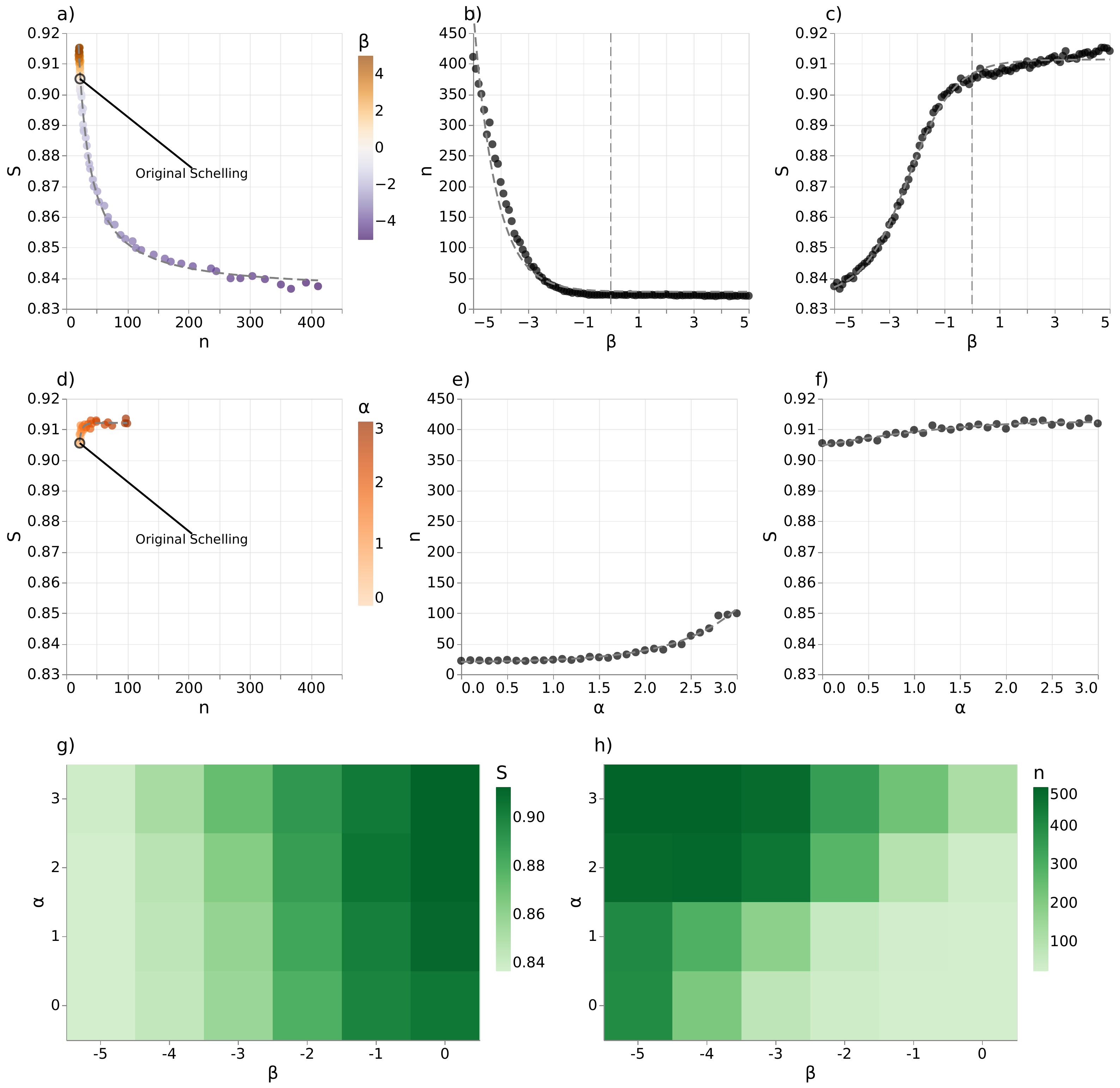}
   \caption{\scriptsize \textbf{Effects of distance and relevance exponents on segregation dynamics with homophily=0.5} 
     (a-c) Effects of $\beta$ on segregation dynamics. (a) The average value of $n$ and $S$ over 100 simulations with the same $\beta$ value but different initial grid configurations, colour-coded by the value of $\beta$. 
    The lower $\beta$, the higher the cost of relocating far away, resulting in longer convergence time and reduced segregation levels compared to the original Schelling model. 
    (b) $\beta$ vs average $n$ over 100 simulations. 
    The lower $\beta$ ($< 0$), the longer the simulation.
    (c) $\beta$ vs average $S$ over 100 simulations.
    For $\beta < 0$, there is an exponential increase in $S$; $\beta > 0$, the growth is moderate.
    (d-f) Effects of $\alpha$ on segregation dynamics. 
    (d) The average value of $n$ and $S$ over 100 simulations with the same value of $\alpha$ but different initial grid configurations, colour-coded by the value of $\alpha$.
    Increasing values of $\alpha$ elongate $n$ and slightly increase $S$. 
    (e) $\alpha$ vs average $n$ over 100 simulations. 
    (f) $\alpha$ vs average $S$ over 100 simulations.
    (g) The average $S$ (colour) for each combination of $\alpha$ and $\beta < 0$. 
    For every value of $\alpha$, higher $\beta$ values lead to a higher $S$; for every $\beta$, higher $\alpha$ values lead to a higher $S$.
    (h) The average $n$ (colour) for each combination of $\alpha$ and $\beta < 0$. 
    For every value of $\alpha$, higher $\beta$ values lead to a lower $n$; for every $\beta$, higher $\alpha$ values lead to higher $n$.
    }
        \label{fig:hom_0.5}

\end{figure}

\begin{figure}[H]
    \centering
\includegraphics[width=.9\textwidth]{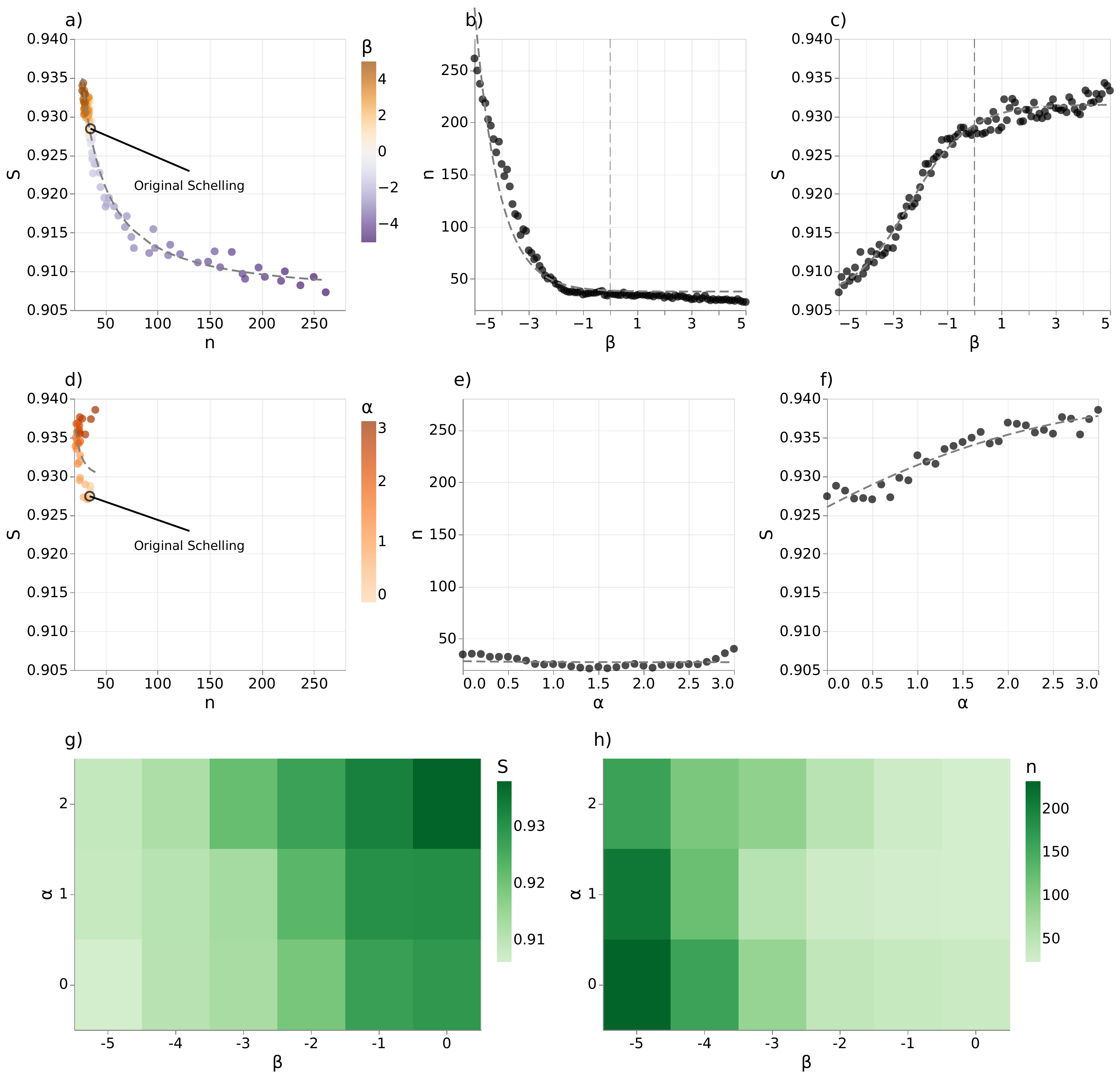}
    \caption{\scriptsize \textbf{Effects of distance and relevance exponents on segregation dynamics with distribution 10\%/90\% and density 0.5} 
     (a-c) Effects of $\beta$ on segregation dynamics. (a) The average value of $n$ and $S$ over 100 simulations with the same $\beta$ value but different initial grid configurations, colour-coded by the value of $\beta$. 
    The lower $\beta$, the higher the cost of relocating far away, resulting in longer convergence time and reduced segregation levels compared to the original Schelling model. 
    (b) $\beta$ vs average $n$ over 100 simulations. 
    The lower $\beta$ ($< 0$), the longer the simulation.
    (c) $\beta$ vs average $S$ over 100 simulations.
    For $\beta < 0$, there is an exponential increase in $S$; $\beta > 0$, the growth is moderate.
    (d-f) Effects of $\alpha$ on segregation dynamics. 
    (d) The average value of $n$ and $S$ over 100 simulations with the same value of $\alpha$ but different initial grid configurations, colour-coded by the value of $\alpha$.
    Increasing values of $\alpha$ elongate $n$ and slightly increase $S$. 
    (e) $\alpha$ vs average $n$ over 100 simulations. 
    (f) $\alpha$ vs average $S$ over 100 simulations.
    (g) The average $S$ (colour) for each combination of $\alpha$ and $\beta < 0$. 
    For every value of $\alpha$, higher $\beta$ values lead to a higher $S$; for every $\beta$, higher $\alpha$ values lead to a higher $S$.
    (h) The average $n$ (colour) for each combination of $\alpha$ and $\beta < 0$. 
    For every value of $\alpha$, higher $\beta$ values lead to a lower $n$; for every $\beta$, higher $\alpha$ values lead to higher $n$.
    }
        \label{fig:pop_10-90}

\end{figure}

    \begin{figure}[H]
    \centering
\includegraphics[width=.9\textwidth]{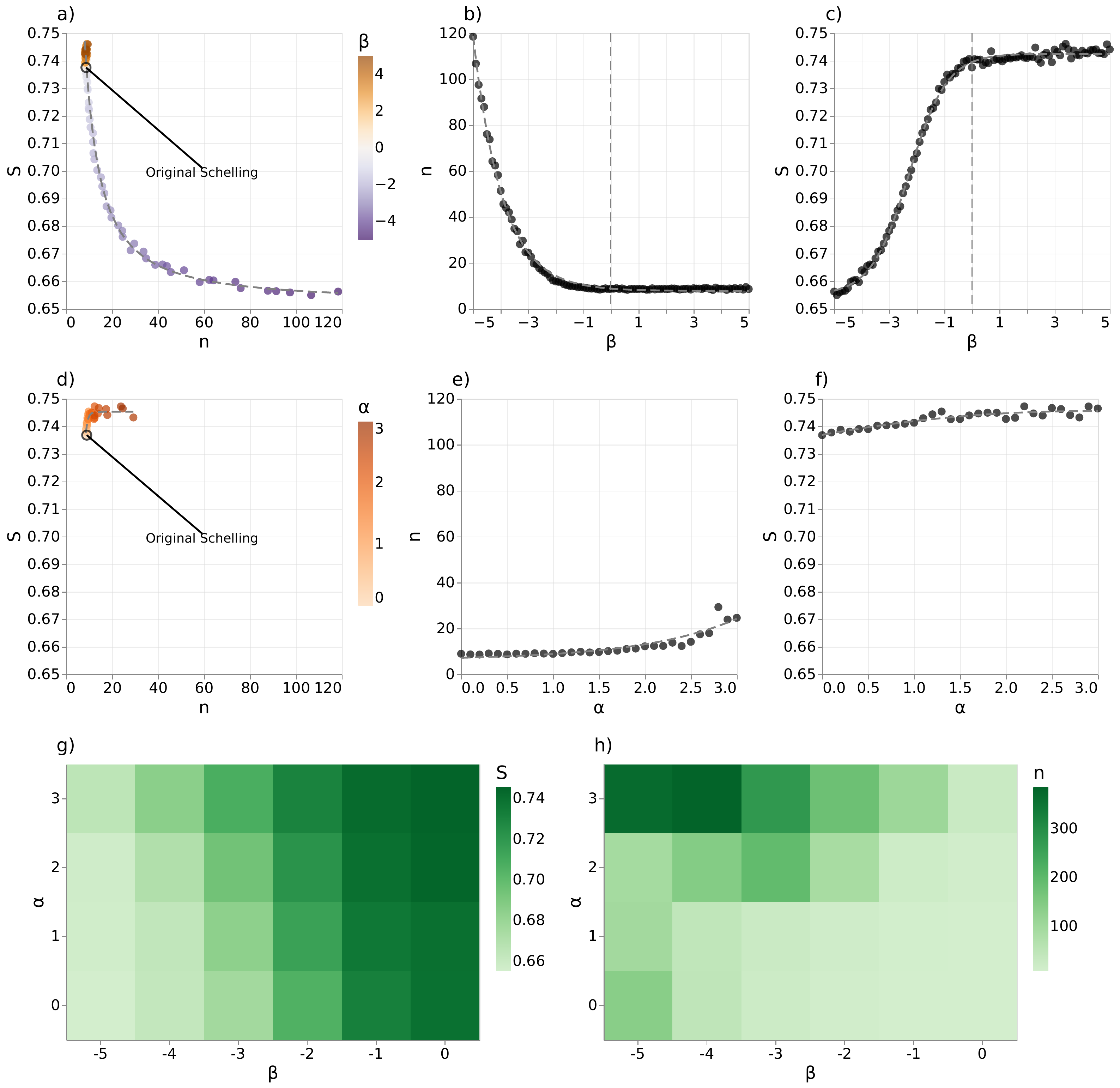}
\caption{\scriptsize \textbf{Effects of distance and relevance exponents on segregation dynamics with distribution of population 50\%/50\%} 
     (a-c) Effects of $\beta$ on segregation dynamics. (a) The average value of $n$ and $S$ over 100 simulations with the same $\beta$ value but different initial grid configurations, colour-coded by the value of $\beta$. 
    The lower $\beta$, the higher the cost of relocating far away, resulting in longer convergence time and reduced segregation levels compared to the original Schelling model. 
    (b) $\beta$ vs average $n$ over 100 simulations. 
    The lower $\beta$ ($< 0$), the longer the simulation.
    (c) $\beta$ vs average $S$ over 100 simulations.
    For $\beta < 0$, there is an exponential increase in $S$; $\beta > 0$, the growth is moderate.
    (d-f) Effects of $\alpha$ on segregation dynamics. 
    (d) The average value of $n$ and $S$ over 100 simulations with the same value of $\alpha$ but different initial grid configurations, colour-coded by the value of $\alpha$.
    Increasing values of $\alpha$ elongate $n$ and slightly increase $S$. 
    (e) $\alpha$ vs average $n$ over 100 simulations. 
    (f) $\alpha$ vs average $S$ over 100 simulations.
    (g) The average $S$ (colour) for each combination of $\alpha$ and $\beta < 0$. 
    For every value of $\alpha$, higher $\beta$ values lead to a higher $S$; for every $\beta$, higher $\alpha$ values lead to a higher $S$.
    (h) The average $n$ (colour) for each combination of $\alpha$ and $\beta < 0$. 
    For every value of $\alpha$, higher $\beta$ values lead to a lower $n$; for every $\beta$, higher $\alpha$ values lead to higher $n$.
    }
        \label{fig:pop_50-50}

    \end{figure}

\begin{figure}[H]
    \centering
\includegraphics[width=.9\textwidth]{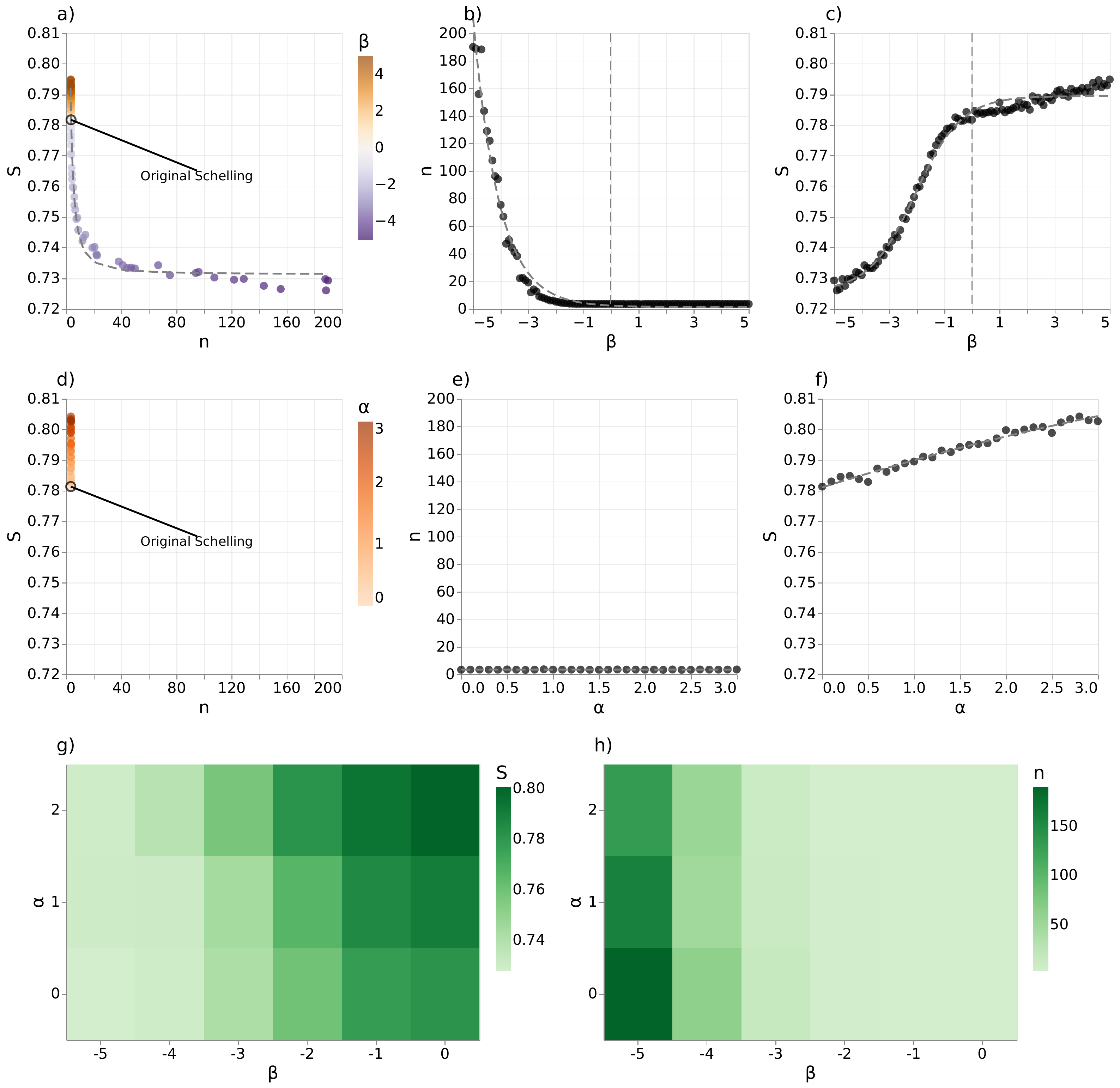}
\caption{\scriptsize \textbf{Effects of distance and relevance exponents on segregation dynamics with Best policy} 
     (a-c) Effects of $\beta$ on segregation dynamics. (a) The average value of $n$ and $S$ over 100 simulations with the same $\beta$ value but different initial grid configurations, colour-coded by the value of $\beta$. 
    The lower $\beta$, the higher the cost of relocating far away, resulting in longer convergence time and reduced segregation levels compared to the original Schelling model. 
    (b) $\beta$ vs average $n$ over 100 simulations. 
    The lower $\beta$ ($< 0$), the longer the simulation.
    (c) $\beta$ vs average $S$ over 100 simulations.
    For $\beta < 0$, there is an exponential increase in $S$; $\beta > 0$, the growth is moderate.
    (d-f) Effects of $\alpha$ on segregation dynamics. 
    (d) The average value of $n$ and $S$ over 100 simulations with the same value of $\alpha$ but different initial grid configurations, colour-coded by the value of $\alpha$.
    Increasing values of $\alpha$ elongate $n$ and slightly increase $S$. 
    (e) $\alpha$ vs average $n$ over 100 simulations. 
    (f) $\alpha$ vs average $S$ over 100 simulations.
    (g) The average $S$ (colour) for each combination of $\alpha$ and $\beta < 0$. 
    For every value of $\alpha$, higher $\beta$ values lead to a higher $S$; for every $\beta$, higher $\alpha$ values lead to a higher $S$.
    (h) The average $n$ (colour) for each combination of $\alpha$ and $\beta < 0$. 
    For every value of $\alpha$, higher $\beta$ values lead to a lower $n$; for every $\beta$, higher $\alpha$ values lead to higher $n$.
    }
        \label{fig:good}

    \end{figure}

\begin{figure}[H]
    \centering
\includegraphics[width=.9\textwidth]{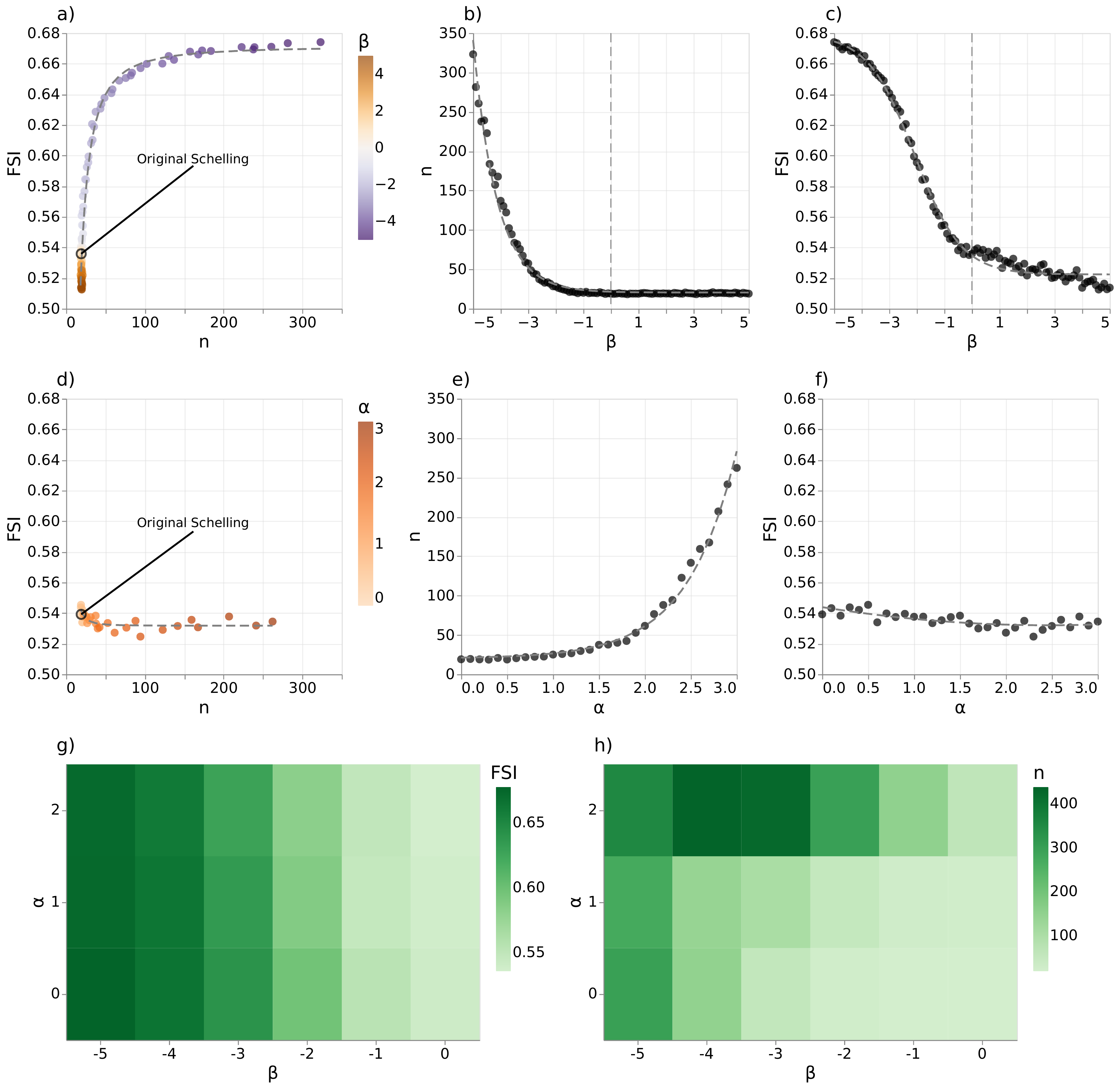}
    \caption{\scriptsize \textbf{Effects of distance and relevance exponents on segregation dynamics using FSI} 
     (a-c) Effects of $\beta$ on segregation dynamics. (a) The average value of $n$ and $FSI$ over 100 simulations with the same $\beta$ value but different initial grid configurations, colour-coded by the value of $\beta$. 
    The lower $\beta$, the higher the cost of relocating far away, resulting in longer convergence time and higher FSI compared to the original Schelling model. 
    (b) $\beta$ vs average $n$ over 100 simulations. 
    The lower $\beta$ ($< 0$), the longer the simulation.
    (c) $\beta$ vs average $FSI$ over 100 simulations.
    For $\beta < 0$, there is an exponential decrease in $FSI$; $\beta > 0$, the growth is more moderate.
    (d-f) Effects of $\alpha$ on segregation dynamics. 
    (d) The average value of $n$ and $FSI$ over 100 simulations with the same value of $\alpha$ but different initial grid configurations, colour-coded by the value of $\alpha$.
    Increasing values of $\alpha$ elongate $n$ and slightly decrease $FSI$. 
    (e) $\alpha$ vs average $n$ over 100 simulations. 
    (f) $\alpha$ vs average $FSI$ over 100 simulations.
    (g) The average $FSI$ (colour) for each combination of $\alpha$ and $\beta < 0$. 
    For every value of $\alpha$, higher $\beta$ values lead to a lower $FSI$; for every $\beta$, higher $\alpha$ values lead to a a slightly lower $FSI$.
    (h) The average $n$ (colour) for each combination of $\alpha$ and $\beta < 0$. 
    For every value of $\alpha$, higher $\beta$ values lead to a lower $n$; for every $\beta$, higher $\alpha$ values lead to higher $n$.
    }
        \label{fig:fsi}

\end{figure}

\begin{figure}[H]
    \centering
\includegraphics[width=.9\textwidth]{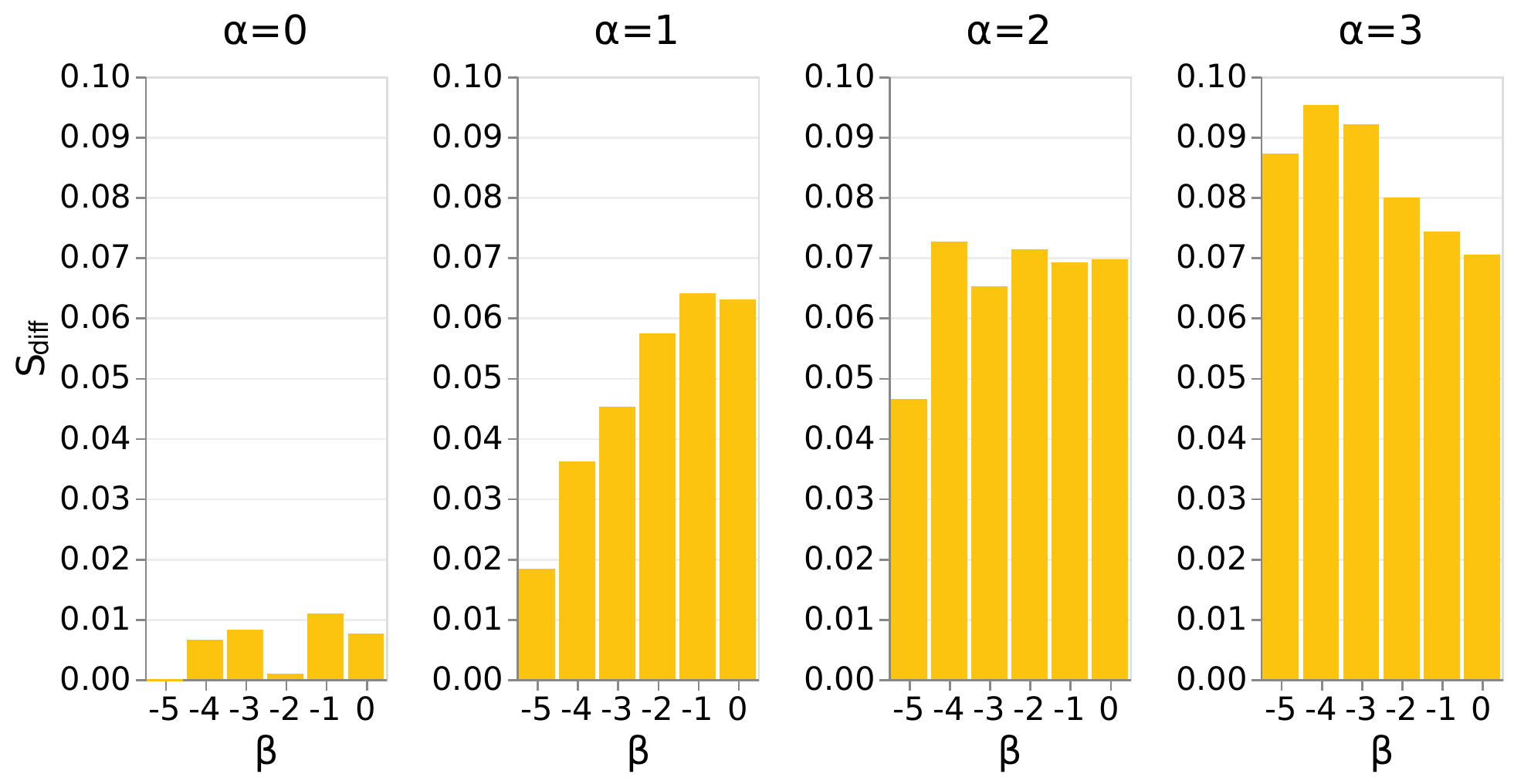}
    \caption{Each plot shows the difference between segregation of the periphery and segregation of center $S_{\tiny \mbox{diff}} = S_{\tiny \mbox{periphery}} - S_{\tiny \mbox{centre}}$ for combinations of values of $\alpha$ and $\beta$. Higher values of $\alpha$ lead to increase of $S_{\tiny \mbox{diff}}$.}
    \label{fig:diff_seg}
\end{figure}

\begin{figure}[H]
    \centering
\includegraphics[width=.9\textwidth]{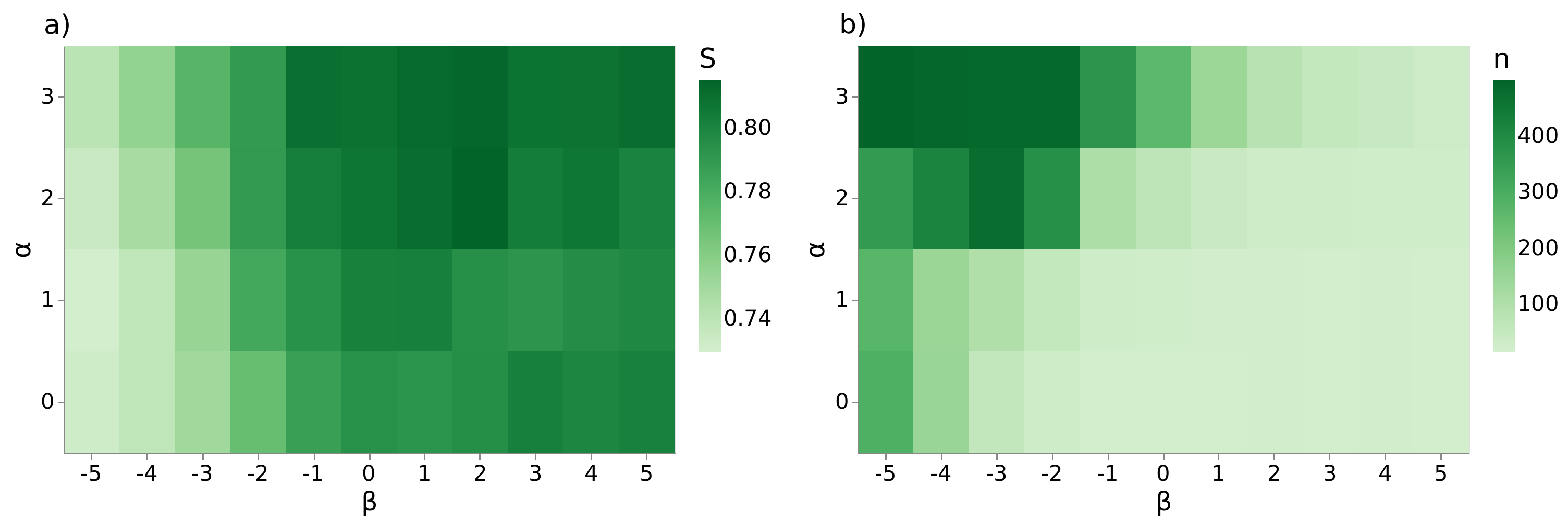}
    \caption{(a) The average $S$ (colour) for each combination of $\alpha$ and $\beta$ also with $\beta > 0$. 
    For positive values of $\beta$, $S$ doesn't show particular evidence of relation with $\alpha$.
    (b) The average $n$ (colour) for each combination of $\alpha$ and $\beta$ also with $\beta > 0$. 
    For positive values of $\beta$, $n$ remains low but doesn't show particular variation among different values of $\alpha$ like in case of $\beta<0$.}
        \label{fig:betapos}

\end{figure}

\begin{figure}[H]
    \centering
\includegraphics[width=.8\textwidth]{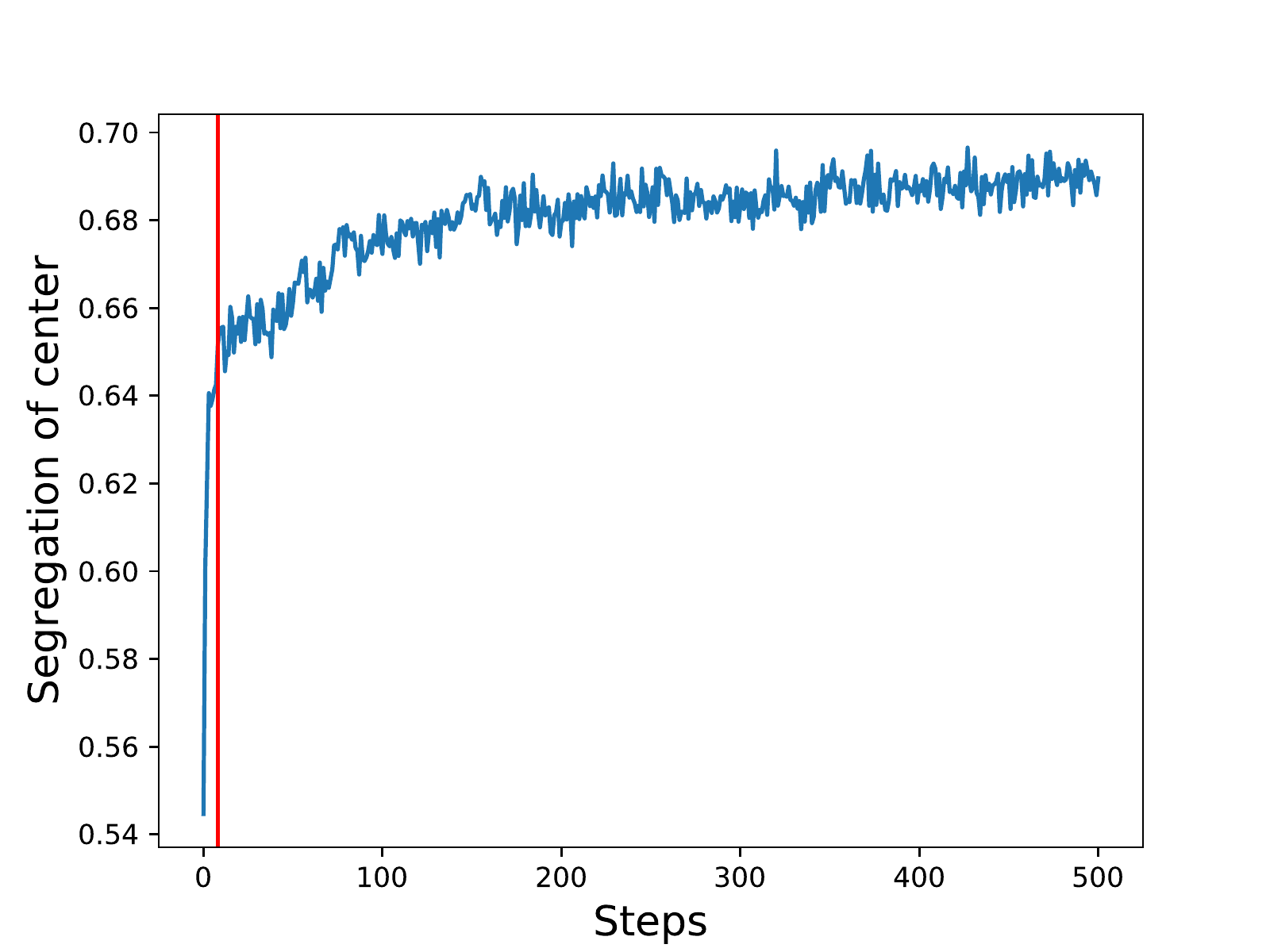}
    \caption{Example of a temporal evolution of the average segregation level within the centre zone during the simulation. The red line indicates the step at which the percentage change calculated respect previous 5 steps is lower than 2\%. }
    \label{fig:tippingpoint}
\end{figure}

\end{document}